\documentclass [structabstract]{aa} % for the abstract without structuration 
\usepackage{subfigure}
\usepackage{longtable,lscape}
\usepackage{rotating}
\usepackage{natbib}
\usepackage{amsmath}
\usepackage{gensymb}
\usepackage{placeins}
\usepackage{afterpage}
\usepackage{tabularx}
\usepackage{booktabs}
\usepackage{multirow}

\usepackage{color}
\usepackage{graphicx}
\usepackage{txfonts}
\usepackage{twoopt}
\usepackage{hyperref}
\usepackage{etoolbox}
\makeatletter
\newcount\c@additionalboxlevel
\newcount\c@maxboxlevel
\patchcmd\@combinedblfloats{\box\@outputbox}{%
  \stepcounter{additionalboxlevel}%
  \box\@outputbox
}{}{\errmessage{\noexpand\@combinedblfloats could not be patched}}

\AtBeginShipout{%
  \ifnum\value{additionalboxlevel}>\value{maxboxlevel}%
    \typeout{Warning: maxboxlevel might be too small, increase to %
      \the\value{additionalboxlevel}%
    }%
  \fi 
  \@whilenum\value{additionalboxlevel}<\value{maxboxlevel}\do{%
    \typeout{* Additional boxing of page `\thepage'}%
    \setbox\AtBeginShipoutBox=\hbox{\copy\AtBeginShipoutBox}%
    \stepcounter{additionalboxlevel}%
  }%
  \setcounter{additionalboxlevel}{0}%
}
\makeatother

\bibpunct{(}{)}{;}{a}{}{,} %% natbib format for A&A and ApJ
\makeatletter
\newcommandtwoopt{\citeads}[3][][]{\href{http://adsabs.harvard.edu/abs/#3}%
{\def\hyper@linkstart##1##2{}%
\let\hyper@linkend\@empty\cite[#1][#2]{#3}}}
\newcommandtwoopt{\citealpads}[3][][]{\href{http://adsabs.harvard.edu/abs/#3}%
{\def\hyper@linkstart##1##2{}%
\let\hyper@linkend\@empty\citealp[#1][#2]{#3}}}
\newcommandtwoopt{\citepads}[3][][]{\href{http://adsabs.harvard.edu/abs/#3}%
{\def\hyper@linkstart##1##2{}%
\let\hyper@linkend\@empty\citep[#1][#2]{#3}}}
\newcommandtwoopt{\citetads}[3][][]{\href{http://adsabs.harvard.edu/abs/#3}%
{\def\hyper@linkstart##1##2{}%
\let\hyper@linkend\@empty\citet[#1][#2]{#3}}}
\newcommandtwoopt{\citeyearads}[3][][]%
{\href{http://adsabs.harvard.edu/abs/#3}
{\def\hyper@linkstart##1##2{}%
\let\hyper@linkend\@empty\citeyear[#1][#2]{#3}}}
\makeatother

\begin{document}

\newcommand{\be}{\begin{equation}}
\newcommand{\ee}{\end{equation}}

\newcommand{\cmc}{cm$^{-3}$}
\newcommand{\cmg}{cm$^{2}$/g}
\newcommand{\kms}{km/s}
\newcommand{\um}{$\mu m$}
\newcommand{\roph}{$\rho$ Oph}
\newcommand{\Msun}{M$_\odot$}
\newcommand{\Lsun}{L$_\odot$}
\newcommand{\Mearth}{M$_\oplus$}
\newcommand{\Ha}{H$\alpha$}
\newcommand{\Pab}{Pa$\beta$}
\newcommand{\Pag}{Pa$\gamma$}
\newcommand{\Brg}{Br$\gamma$}
\newcommand{\Teff}{T$_{eff}$}
\newcommand{\Lstar}{L$_\ast$}
\newcommand{\Rstar}{R$_\ast$}
\newcommand{\Mstar}{M$_\ast$}
\newcommand{\Mdisk}{M$_{disk}$}
\newcommand{\Mdust}{M$_{dust}$}
\newcommand{\Lacc}{L$_{acc}$}
\newcommand{\Lx}{L$_{\rm X}$}
\newcommand{\Macc}{$\dot M_{acc}$}
\newcommand{\Mloss}{$\dot M_{loss}$}
\newcommand{\Mwind}{$\dot M_{wind}$}
\newcommand{\Myr}{M$_\odot$/yr}
\newcommand {\Fmm} {F$_{890 \mu m}$}
\newcommand {\opmm} {$\kappa_{890 \mu m}$}
\newcommand {\MJ} {M$_{\rm J}$}
\newcommand{\MOP}{M$_{disk}\times \kappa_{890 \mu m}$}
\newcommand{\twom}{2M0444$+$2512}
\newcommand{\cida}{CIDA~1}
\newcommand{\ctfour}{CFHT~Tau4}

\newcommand{\simless}{\mathbin{\lower 3pt\hbox
      {$\rlap{\raise 5pt\hbox{$\char'074$}}\mathchar"7218$}}}
\newcommand{\simgreat}{\mathbin{\lower 3pt\hbox
     {$\rlap{\raise 5pt\hbox{$\char'076$}}\mathchar"7218$}}}

   \title{Brown dwarf disks with ALMA:\\Evidence for truncated dust disks in Ophiuchus
   %Oph BDs with ALMA
   %\thanks{}
   }

   \author{
          L. Testi\inst{1,2,3,4},  A. Natta\inst{5,3},  A. Scholz\inst{6}, 
          M. Tazzari\inst{1,2}, L. Ricci\inst{7}, I. de Gregorio Monsalvo\inst{1,8}
          }

 \institute  {ESO/European Southern Observatory, Karl-Schwarzschild-Strasse 2 D-85748 Garching bei M\"unchen, Germany
              \email{ltesti@eso.org} 
\and 
  {Excellence Cluster ``Universe'', Boltzmann str. 2, D-85748 Garching bei Muenchen, Germany}
\and 
  {INAF/Osservatorio Astrofisico of Arcetri, Largo E. Fermi, 5, 50125 Firenze, Italy}
  \and 
  {Gothenburg Center for Advance Studies in Science and Technology,
Department of Mathematical Sciences, Chalmers University of Technology and University of Gothenburg, SE-412 96 Gothenburg, Sweden}
        \and
   {School of Cosmic Physics, Dublin Institute for Advanced Studies, 31 Fitzwilliams Place, Dublin 2, Ireland}
   \and 
              SUPA, School of Physics and Astronomy, University of St. Andrews, North Haugh, St. Andrews, Fife KY16 9SS, United Kingdom
\and 
            Harvard-Smithsonian Center for Astrophysics, 60 Garden Street, Cambridge, MA 02138, USA   \and 
              Joint ALMA Observatory (JAO), Alonso de Cordova 3107 Vitacura -Santiago de Chile
%\and 
% {INAF/Osservatorio Astronomico di Capodimonte, Salita Moiariello, 16  80131, Napoli, Italy}
%         \and
%   {Department of Planetary Science, Lunar and Planetary Lab, University of Arizona, 1629, E. University Blvd, 85719, Tucson, AZ, USA}
%       \and 
%   {INAF/Osservatorio Astronomico di Palermo, Piazza del Parlamento 1, 90134 Palermo, Italy }
%        \and
% {ASI-Science Data Center, Via del Politecnico snc, I-00133 Rome, Italy }\\
}
\authorrunning{Testi et al.}
\titlerunning{Brown dwarf disks in Ophiuchus}
        \offprints{ltesti@eso.org}
   \date{Received ...; accepted ...}

 \abstract 
 {The study of the properties of disks around young brown dwarfs can provide important clues on the formation of these very low-mass objects and on the possibility of forming planetary systems around them. The presence of warm dusty disks around brown dwarfs is well known, based on near- and mid-infrared studies. }
 {High angular resolution observations of the cold outer disk are limited; we used ALMA
to attempt a first survey of young brown dwarfs in the \roph\ star-forming region.}
 {All 17 young brown dwarfs in our sample were observed at 890~$\mu $m in the continuum at $\sim0.\!^{\prime\prime}5$ angular resolution. The sensitivity of our observations was chosen to detect $\sim0.5$ \Mearth\ of dust.}
 {We detect continuum emission in 11 disks ($\sim65$\%\ of the total), and the estimated mass of dust in the detected disks ranges from $\sim0.5$ to $\sim6$~\Mearth. These disk masses imply that planet formation around brown dwarfs may be relatively rare  and that the supra-Jupiter mass companions found around some brown dwarfs are probably the result of a binary system formation. We find evidence that the two brightest disks in \roph\ have sharp outer edges at R$\simless$25~AU, in contrast to disks around Taurus brown dwarfs. This difference may suggest that the different environment in \roph\ may lead to significant differences in disk properties. A comparison of the \Mdisk/\Mstar\ ratio for brown dwarf and solar-mass systems also shows a possible deficit of mass in brown dwarfs, which could support the evidence for dynamical truncation of disks in the substellar regime. These findings are still tentative and need to be put on firmer grounds by studying the gaseous disks around brown dwarfs and by performing a more systematic and unbiased survey of the disk population around the more massive stars.}{} 
 
% 5 {} token are mandatory

 \maketitle
%
%________________________________________________________________

\section{Introduction}\label{introduction}

After decades since the discovery of the first young brown dwarf \citepads[BDs;][]{%1995Natur.378..463N,
1995Natur.377..129R} and their disks \citepads{1998A&A...335..522C,2001A&A...376L..22N,2002A&A...393..597N}, the formation mechanism of very low-mass objects (below $\sim 0.1$ \Msun) is still  controversial.  Possibilites range from the collapse of a single low-mass core \citepads{2009ApJ...702.1428H}, as for the more massive T Tauri stars (TTS),  to ejection of low-mass protostellar seeds from multiple stellar systems before accretion terminates  \citepads{2001AJ....122..432R,2004Ap&SS.292..297B,2009MNRAS.392..590B} or from fragmenting circumstellar disks \citepads{2009MNRAS.392..413S}. A solar-mass core in the vicinity of an OB star may be photo-eroded to become a very low-mass object by the time of collapse \citepads{1996AJ....111.2349H,2004A&A...427..299W}. A variety of formation channels may in fact co-exist and their relative importance may change as a function of mass and the environment.  All the formation models account for the presence of circumstellar accretion disks around young BDs, but the current generation of numerical models is still unable to reliably predict how much mass is retained in the small disks around BDs when they are ejected.
 A solid expectation is that the ratio of the disk mass over the mass of the central object should drop significantly if the ejection process becomes an important formation channel \citepads{2015MNRAS.449.3432S}. A reliable observational estimate of the (possible) variation of this ratio with the mass of the central object would be a critical test that models would have to compare against. 

Knowledge of the disk mass is also important to assess the potential of planet formation around BDs,  and, in particular the formation of planetesimals and rocky planets. 
Infrared spectroscopy of young BDs with disks has shown that in several sources dust processing similar to that observed in more massive disks around young stars has taken place in the disks atmospheres \citepads{2004A&A...426L..53A,2005Sci...310..834A,2004A&A...427..245S}.  Growth to millimeter and centimeters sizes on the disk midplane, similar to what occurs in TTS disks,  has been inferred for few BDs observed with ALMA and CARMA \citepads{2012ApJ...761L..20R,2013ApJ...764L..27R,2014ApJ...791...20R}. All this
suggests that the initial steps of the core accretion scenario can indeed occur in BD disks, however, 
theoretical models of grain growth \citepads{2013A&A...554A..95P} in  low-mass disks  encounter very serious difficulties; this challenges  current views of how  and when grain growth may occur.  \citetads{2007MNRAS.381.1597P} estimate that the disks around BDs need to be relatively massive, at least a few Jupiter masses, to enable  efficient terrestrial planet formation.

Measuring the mass distribution of disks around a significant sample of young BDs is thus  essential to shed light on the BD formation mechanisms and to
assess their potential to form planetary systems.  In this paper we present the first ALMA observations of an unbiased  sample of very low-mass objects in the very young star-forming region \roph.

\section {The sample}
\label{s_sample}

   \begin{table*}
%   \centering
\begin{center}
      \caption[]{Sample }
         \label{table_sample}
    \begin{tabular}{l c c c c c c c c  l}
        \hline
Name&                      $\alpha$  &        $\delta$ &      ST&    T$_{eff}$&  A$_V$&  L$_\ast $ & M$_\ast $&  Ref& Other Names\\
         &                          (2000)             &    (2000)                 &           &    [K]          &  [mag]&   [ L$_\odot $]&   [M$_\odot $]&  &  \\
\hline
SONYC-RhoOph-8 & 16:26:18.58&  -24.29.51.8 & M7&  2900&  18.8&  0.1   &   0.1&   2 & CFHTWIR-Oph16\\
ISO-Oph023  &16:26:18.81&  -24.26.10.8 &  M7  &  2900&    9.7& 0.04  &  0.07 &  1& CRBR 2317.3-1925 \\
ISO-Oph030&  16:26:21.53&   -24.26.01.4&  M7 &  2900  &  4.5  &0.07 & 0.1  & 1& GY92-5\\
ISO-Oph032               &16:26:21.90&    -24.44.40.09&  M6.5&  2935 & 0.6 & 0.03& 0.07  & 1 & GY92-3\\
ISO-Oph033  &                    16:26:22.26&      -24.24.07.55&    M8   &  2700&  7.7& 0.005&  0.03 & 1 & GY92-11\\
ISO-Oph035  &   16:26:22.96   &   -24:28:46.1  &        M6 & 3000&  10.7    & 0.06 &0.1   &2  &GY15, SONYC-RhoOph-5\\
CRBR 2322.3-1143 &    16:26:23.81   &   -24:18:29.0 &    M6.5 &2935  &8.6 & 0.01 & 0.07 & 2  & \\  
ISO-Oph042 &                    16:26:27.80  &   -24.26.42.22&   M5&  3100 &6.2 & 0.03 & 0.1& 3   & GY92-37\\
GY92-202  &                  16:27:05.98   &     -24:28:36.3  &     M7& 2900 &13.0  & 0.02& 0.06  &   2 & \\
ISO-Oph102  &                16:27:06.58&  -24.41.49.28&     M5 &3125 &2.2&    0.047& 0.15 &  1  & GY92-204\\
ISO-Oph138&           16:27:26.22&  -24.19.23.46&    M7.75&  2753 & 16.4&  0.1 & 0.05& 2   & SONYC-RhoOph-10, CFHTWIR-Oph78 \\ 
GY92-264 &                16:27:26.57&   -24.25.54.77&   M8& 2700& 2.2 &  0.023& 0.03 &   4    &  \\
ISO-Oph160 &      16:27:37.42&        -24.17.55.34&     M7.5 &2800& 6.1& 0.03& 0.06&1  & \\
ISO-Oph164 &           16:27:38.63&      -24:38:39.19 &     M8  &2700&  5.1&0.05  & 0.05& 1&GY310, ROXN62\\
GY92-320    &                    16:27:40.84&         -24:29:00.8&   M7.75 &  2753& 2.4&  0.007& 0.04& 2 &CFHTWIR-Oph 96\\ 
ISO-Oph176  &       16:27:46.29 &        -24:31:41.2 &          M7.5& 2800 &6.9 & 0.06 & 0.06 & 1 & GY92-350\\ 
ISO-Oph193 &                                       16:28:12.7&     -24.11.36.08&    M6&  3000 &7.4& 0.07 &0.11 &1 &  B162812-241138\\  
 \end{tabular}
\end{center}
References to stellar parameters: 1: Manara et al. (2015); 2: data from Alves de Oliveira et al. (2012) and references therein (see text); 3: data from Muzic et al. (2012) (see text); 4: Liu et al. (2015).
   \end{table*}
   
The sample includes 17  spectrally confirmed \roph\  BDs with infrared excess, which indicates the presence of a circumstellar disk.  
The sample includes about half of all the substellar objects with infrared excess confirmed to date in \roph\
\citepads{2002A&A...393..597N,2002ApJ...571L.155T,2011ApJ...726...23G,2012A&A...539A.151A,2012ApJ...744..134M}
At the time of the proposal (July 2012, Cycle I), they were all the spectroscopically confirmed Class II BDs in \roph.  No selection based on Herschel fluxes or pre-existing (sub)mm observations was applied. 
%In the meantime,  new spectral surveys have identified a number of additional BDs (Geers et al. 2011, Alves de Oliveira et al. 2012, Muzic et al. 2012), bringing the number of Class II BDs to about 30.  
It should be noted that an object 
in \roph\ is considered a BD if it has spectral type of M6 or later. Nonetheless, given the difficulties in estimating spectral types in highly extincted objects, some objects may or may not be included in the published BD samples, depending on the adopted spectral type determination. Recently, the stellar parameters of a subset (eight objects) of our sample have been re-determined using X-Shooter spectra by \citetads{2015A&A...579A..66M}, who also
 provide a detailed discussion of the uncertainties.  In two cases, the spectral type has been revised to M5, which is just above the BDs threshold. We kept the objects in our analysis, even if strictly speaking they do not meet our original selection criterion.

In spite of the continuing effort to identify and classify new objects, it is important to stress that we are always dealing with  extinction-limited samples, and not with samples complete down to a given mass sensitivity.  Moreover, as mentioned above, the determination of photospheric parameters of  \roph\ objects (for any mass) is very uncertain because of the large extinction,  young ages, and uncertainties in the evolutionary tracks \citepads[see][]{2015A&A...579A..66M}.

Table \ref{table_sample} summarizes the adopted values of the stellar parameters  for our sample objects.  For those not in \citetads{2015A&A...579A..66M}, we  estimated the luminosity from the ST, extinction, and J magnitude provided by \citetads{2012A&A...539A.151A} and \citetads{2012ApJ...744..134M}, the effective temperature using the ST-temperature scale of Luhman et al.(2003),  and the mass by comparison with \citetads{2008A&A...482..315B} tracks. The objects range 
in luminosity between $\sim 0.01$ and 0.1 \Lsun\ and   between $\sim 0.03$ and 0.1 \Msun\  in mass.

\section {Observations and results}

\begin{figure*}
        \begin{center}
                \includegraphics[width=18cm]{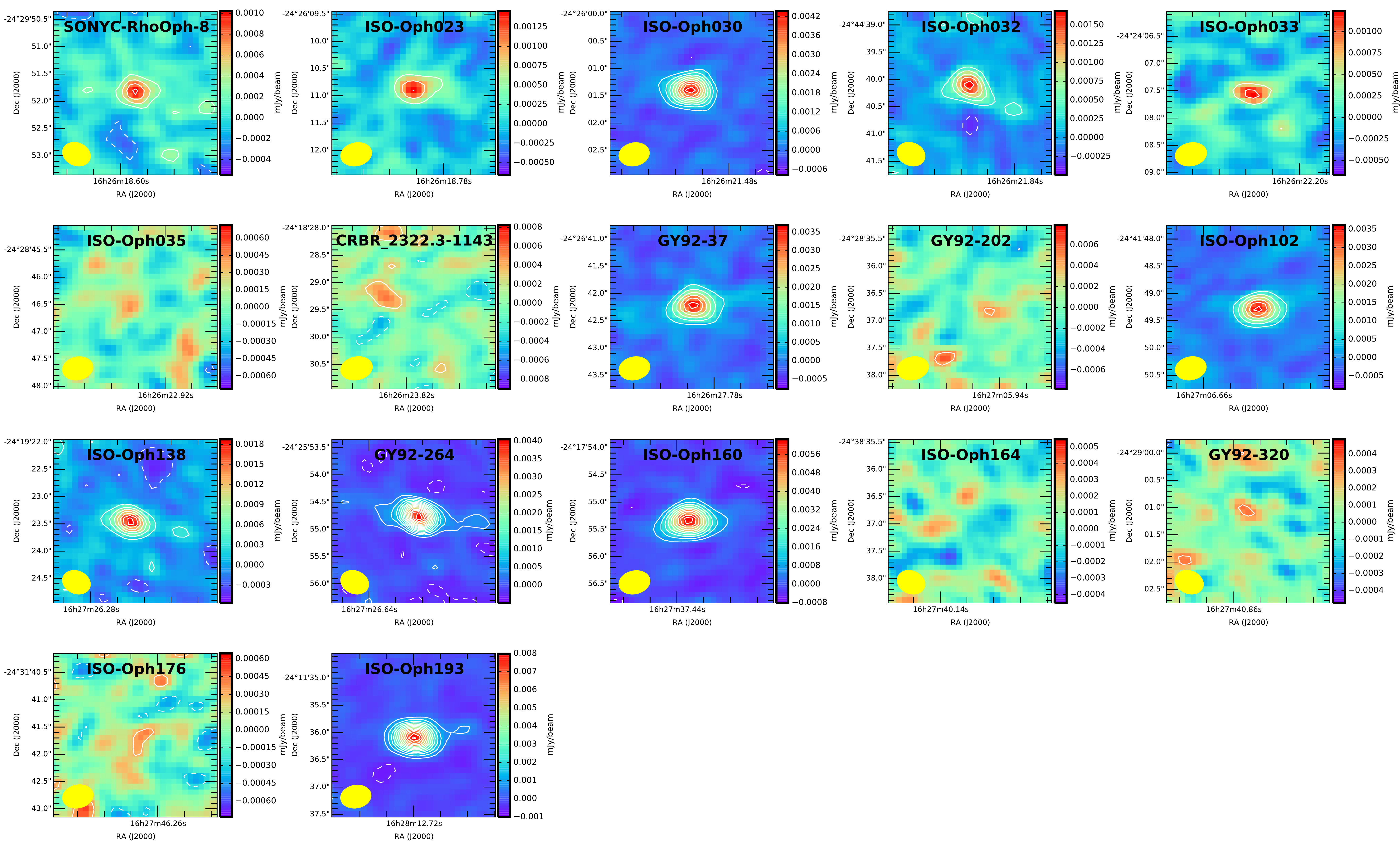}
        \end{center}
        \caption{ALMA Band~7 images of all the observed sources, as labelled in each panel; each panel is $3^{\prime\prime}\times 3^{\prime\prime}$ in size. The synthesized beam full width at half maximum is shown in the lower left corner of each panel; the typical size is $0.\!^{\prime\prime}5\times 0.\!^{\prime\prime}4$. In each panel, contour levels are plot at -3$\sigma$ (dashed), and every 3$\sigma$ from 3$\sigma$ to the maximum.}
\label{f_images}
\end{figure*}

ALMA observed our sources on April 25 and 26, 2014 with 35 and 36 antennas, respectively.
We used the ALMA Band~7 receivers tuned to a frequency of $\sim$338~GHz, the total
effective bandwidth usable for continuum was approximately 6~GHz and the $^{13}$CO(3-2) line was covered in one of the spectral windows of the correlator.
Titan was used as flux calibrator, while J1626$-$2951 and J1625$-$2527 were
used as complex gain calibrators in both sessions. The passband calibrator
was J1733$-$1304 for the first session and J1517$-$2422 for the second session.
Standard calibration was performed by the ESO ALMA Regional Centre,
the flux density scale is expected to be accurate within 5\%.
The total time on source for our targets was three~minutes for the majority of the 
sources; the on-source time was six~minutes for six objects: SONYC-RhoOph-8, 
ISO-Oph032, ISO-Oph138, GY92-264, ISO-Oph164, and GY92-320.

Imaging was performed using natural weighting of the visibilities 
(see Figure~\ref{f_images}). None of
the sources were sufficiently bright to attempt self-calibration. 
In Table~\ref{table_alma} we report in column 2  the peak flux and rms
measured on the images. The full width at half maximum of the synthesized beam
is approximately $0.\!^{\prime\prime}5\times 0.\!^{\prime\prime}4$ for all maps.
$^{13}$CO(3-2) emission was not detected in any of the sources.

The total fluxes reported in column 3 of Table~\ref{table_alma} for the 
detected sources are computed using the CASA
task {\tt uvmodelfit}: we fitted Gaussian sources to the visibilities 
of all detected sources except for the three fainter ones, for which we used 
a point source. The flux upper limits for the non-detected sources correspond
to 98\%\ confidence levels and are computed as three times the rms level above
the measured flux (assumed null when negative) at the nominal position of the object
(see Barenfeld et al. 2016, submitted). For three objects the Gaussian fits imply
a deconvolved source size with a major axis greater or equal than $\sim$0.2~arcsec (see
column 4 of Table~\ref{table_alma}).

%The observed 890 \um\ peak and total flux are given in table \ref{table_alma}.  
%We have added to our sample the object ISO-Oph102, previously observed with Alma by Ricci et al. (2013). Of the 17 observed objects, 11 are detected, with fluxes of 1--10 mJy,  while for the remaining 6 we obtained  upper limits. The observed fluxes are plotted in Fig.~\ref{fig_flux}
%as a function of the stellar luminosity. Arrows indicate 3$\sigma$ upper limits. There is no obvious trend of \Fmm\ with \Lstar.

%Two objects (names) are resolved even at the relatively low resolution of these observations; the corresponding sizes are ?? {\bf Leo: units?}.

   \begin{table*}
      \caption[]{Alma data and disk mass}
         \label{table_alma}
\begin{center}
    \begin{tabular}{l c c  c   c c}
        \hline
Name &  F$_{peak}$ &  F$_{tot}$ & Size$^a$& M$_{dust}^b$&M$_{disk}^c$ \\
          &   [mJy/beam]&    [mJy]&     [arcsec] & [M$_\oplus $]& [$10^{-3}$M$_\odot$] \\
\hline

SONYC-RhoOph-8 &    1.00$\pm$0.08&   1.0&              $<$ 0.2& 0.5& 0.15 \\
ISO-Oph023&    1.30$\pm$0.16  &  1.5&  $<$ 0.2 & 1.1&0.33\\
ISO-Oph030                      &    4.20$\pm$0.15&     4.8 &     $<$0.2&2.7 &0.80\\
ISO-Oph032 &                       1.62$\pm$ 0.1  & 1.80  &   $<$0.2&1.4& 0.42\\  
ISO-Oph033 &                     1.2$\pm$0.18  &    1.3:   & $<$0.2  &2.6& 0.77\\
ISO-Oph035&    -- &         < 0.9 &   --&< 0.5& < 0.15\\
CRBR 2322.3-1143 &  --      &     < 0.6  &  --&< 0.9 & < 0.27\\
ISO-Oph042 &                        3.61$\pm$0.19&  4.2  & $\sim$0.25&3.5& 1.0\\
GY92-202&                         -- &          <0.5 &     --&< 0.5& < 0.15\\
ISO-Oph102 &                     3.66$\pm$0.19 &   3.8  &       $<$0.2&2.5& 0.75\\
ISO-Oph138&  1.96$\pm$0.085 &2.0 &      $<$0.2 &1.0& 0.30\\
GY92-264 &                           3.96$\pm$0.09 & 4.1  &$<$0.2&4.4& 1.3\\
ISO-Oph160&            6.13$\pm$0.20&  7.60  &   $\sim$0.3&6.3& 1.9\\
ISO-Oph164 &                -- &       < 0.8  &    --&< 0.5& < 0.15\\
GY92-320&       --&     <0.4& --&< 0.7& < 0.21 \\
ISO-Oph176&          --    & < 0.5&  --& <0.3& < 0.09\\
ISO-Oph193&    7.82$\pm$0.20 &8.7& $\sim$0.2& 4.8& 1.43 \\
        \end{tabular}
\end{center}
Notes: upper limits are 3$\sigma$ (change). $^a$: Deconvolved major axis; unresolved objects have size$<0.2$ arcsec.
$^b$ Dust mass from the \Fmm, \opmm=2\cmg; see text for details.
$c$ Total (gas+dust) disk mass computed from \Mdust\ assuming a gas=to-dust mass ratio of 100.
   \end{table*}

%\begin{figure}
%       \begin{center}
%               \includegraphics[width=9cm]{flux_lstar_almabd}
%       \end{center}
%       \caption{Observed flux at 890 \um\ as function of the object luminosity \Lstar. Arrow are 3$\sigma$ upper limits.}
%\label{fig_flux}
%\end{figure}

\section {Disk properties}

\subsection{Disk masses}

We determine the mass of dust in the disk  from the 890 \um\ flux using the approximation that the emission is optically thin and isothermal at temperature $T$,
\begin{equation}
\label{eq_1}
F_\nu= {{1}\over{D^2}} \> \kappa_\nu B_\nu(T) \> M_{dust} 
,\end{equation}
where $D$ is the distance and $\kappa_\nu$ the opacity at the frequency of the observation.
This expression has been extensively used in the literature and provides reliable estimates of the product \Mdust$\times \kappa_\nu$ from sub-mm and mm fluxes if the emission is optically thin. This is indeed the case for all the BDs studied in detail so far (see
\citetads{2012ApJ...761L..20R,2013ApJ...764L..27R,2014ApJ...791...20R} and Sec.4.2). In the following, we adopt a dust opacity 
\opmm =2 cm$^2$/g, as in \citetads{2014ApJ...791...20R},  and a temperature that depends on the stellar luminosity as
$T$=25 (\Lstar/\Lsun$)^{0.25}$K
($T_A$ in the following), as suggested by \citetads{2013ApJ...771..129A} in their study of more luminous objects (\Lstar $\sim 0.1-100$ \Lsun). Recently, the choice of the temperature appropriate to recover the dust mass from a single-wavelength (sub)mm flux in lower-luminosity objects was discussed by \citetads{2016ApJ...819..102V} and Daemgen et al. (2016, A\&A submitted). Our motivations for choosing $T_A$ and the uncertainties on \Mdust\ due to the uncertainty on $T$ in eq.(1) (less than a factor 2 in all our objects) are summarized in Appendix~\ref{app_t}.

Of the 17  \roph\ BDs observed with ALMA, 11  have a measured dust mass of 0.5--6 \Mearth\ and 6 have upper limits of $\simless 1$ \Mearth. Assuming a gas-to-dust mass ratio of 100, the total disk mass (gas+dust) is \Mdisk$\sim 1.5-9 \times 10^{-4}$ \Msun.
We do not find more massive disks in our sample and the ratio of the disk-to-star mass is above $\sim$1\%\ only in 4 objects.  
Our results are more accurate, but they are in broad agreement with the disk masses derived by \citetads{2013A&A...559A.126A} from 
  an Herschel PACS survey of BDs in \roph. For 12 objects, they were able to determine  the disk properties   by  fitting the observed SEDs  with parametric disk models. The mass range they derive is of 0.1-30 \Mearth . In particular, when the choice of  the mm opacity, which is usually not constrained by the Herschel data, is taken into account,  masses for 5 of the 6 objects in common  agree within a factor of about $\pm$3;
  % one object we do not detect (ISO-Oph164), and 
 in one case (ISO-Oph042, also known as GY92-264) we find a mass a factor of $\sim 10$ larger, but the uncertainties quoted for this object by \citetads{2013A&A...559A.126A} span more than two orders of magnitude (and include our value).

%\begin{figure}
%       \begin{center}
%               \includegraphics[width=9cm]{mdisk_mstar_almabd}
%       \end{center}
%       \caption{Disk masses  as function of the central object mass \Mstar. Arrow are 3$\sigma$ upper limits. Disk masses have been computed from the 890 \um\ flux assuming a temperature T=25K((\Lstar/\Lsun)**0.25,  a 890 \um\ opacity of 2 cm$^2$/g}. 
%\label{fig_almabd}
%\end{figure}

\subsection {Disk sizes and surface density profile}
\label{s_rad}
   
\begin{figure}
        %\begin{center}
  \includegraphics[width=4.3cm]{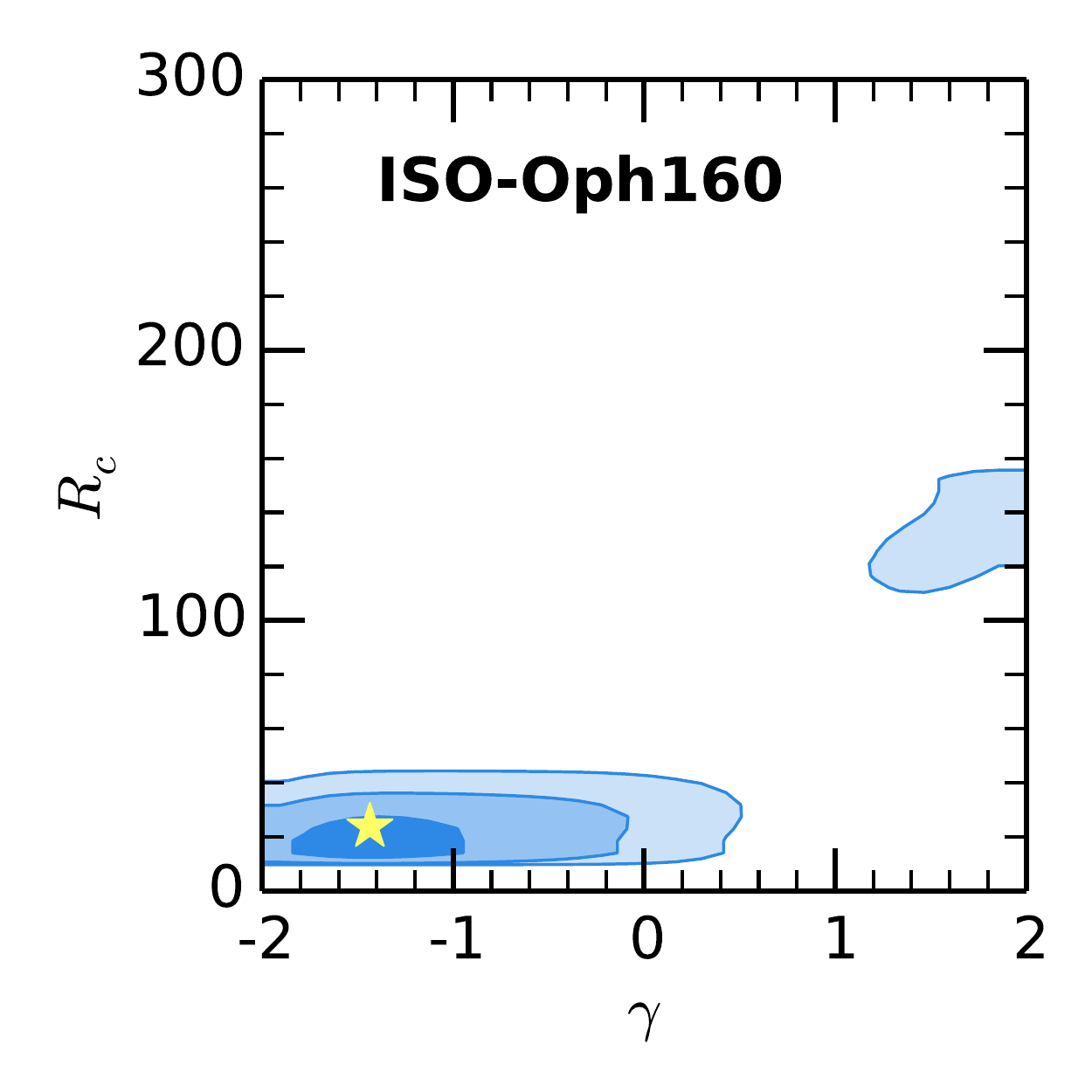}
  \includegraphics[width=4.3cm]{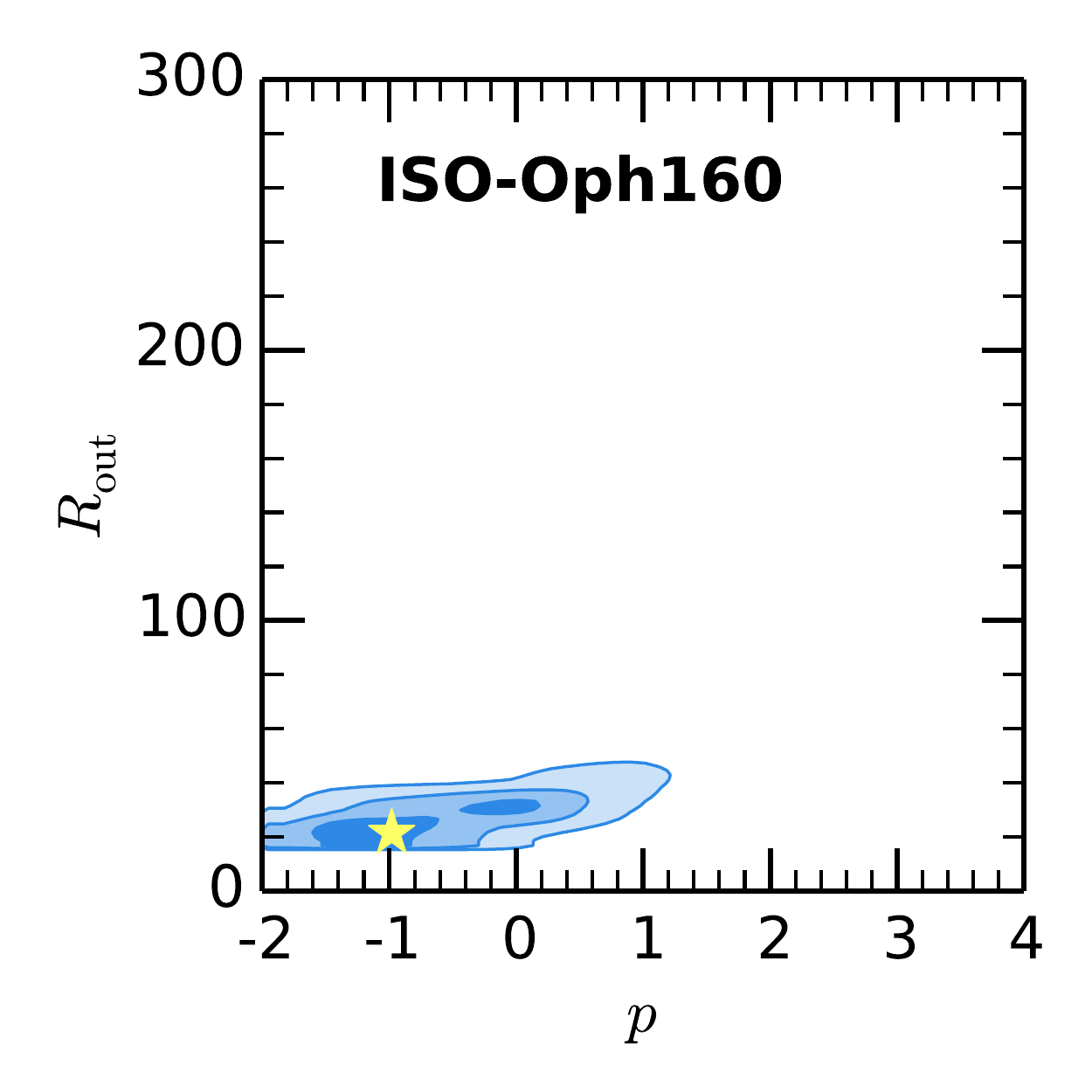}  
  \\
  \includegraphics[width=4.3cm]{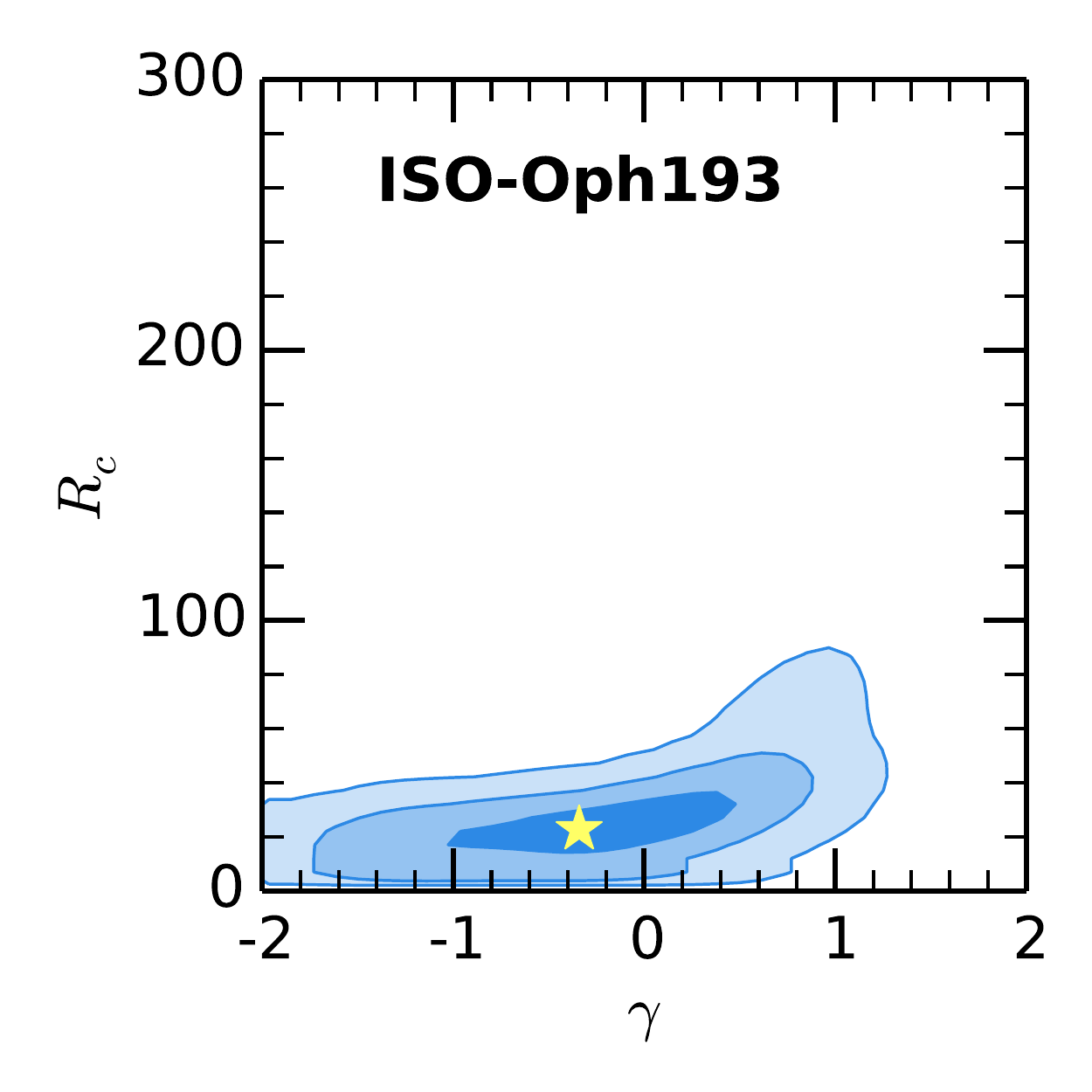}\includegraphics[width=4.3cm]{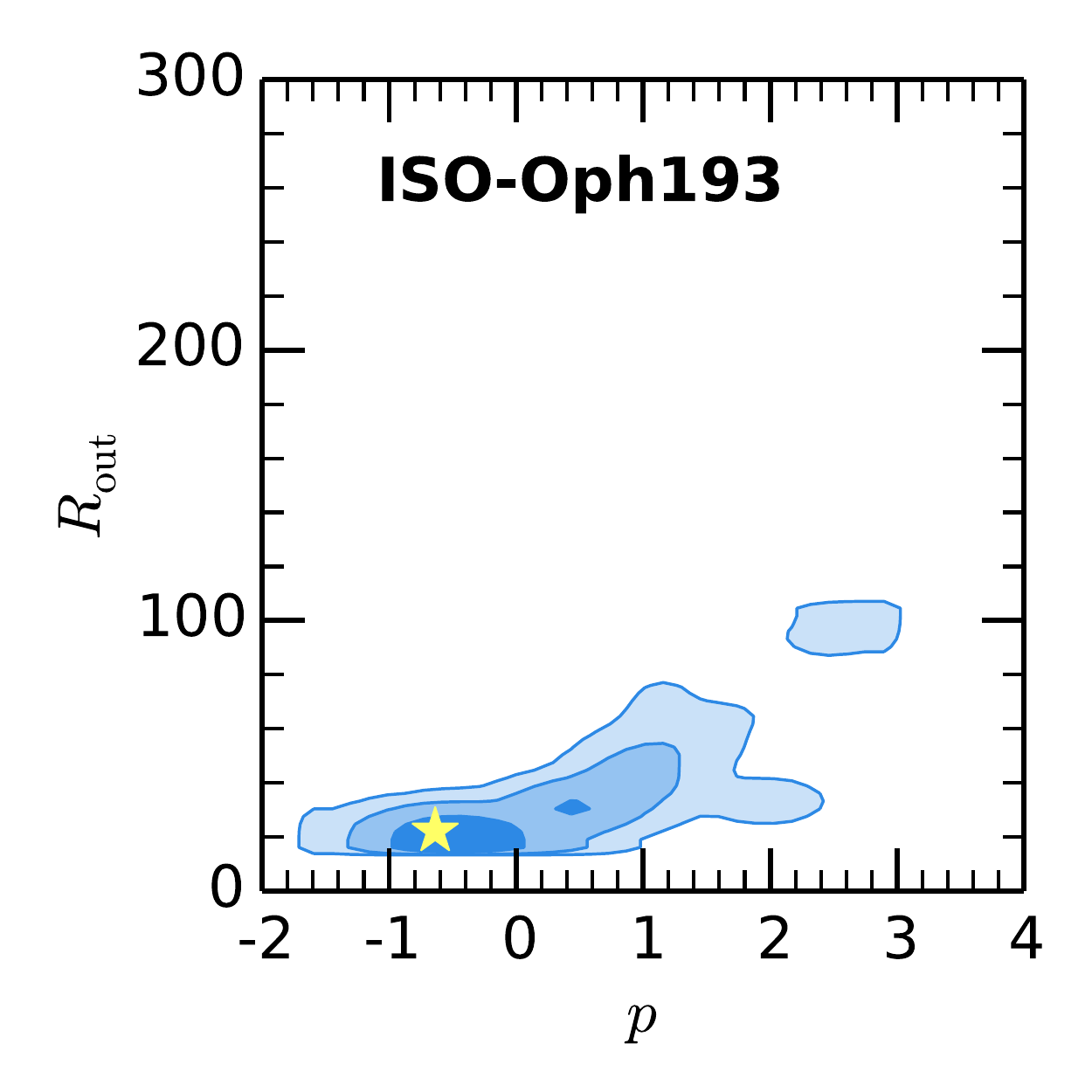}
        %\end{center}
        \caption{Two-dimensional distributions of the model parameters \{$R_c$,$\gamma$\}\ (left column) and \{$R_{out}$,$p$\}\ (right column) for the fits of the disks in ISO-Oph160 (top row) and ISO-Oph193 (bottom row). The different shaded areas correspond to the 1, 2, and 3$\sigma$ confidence levels (from light to dark blue). The yellow star indicates the values of the parameters for the best-fit models shown in Fig.~\ref{f_fituvp} and~\ref{f_fitres}.}
     \label{f_fitpar}
\end{figure}
 
   \begin{figure}
        %\begin{center}
                \includegraphics[width=7cm]{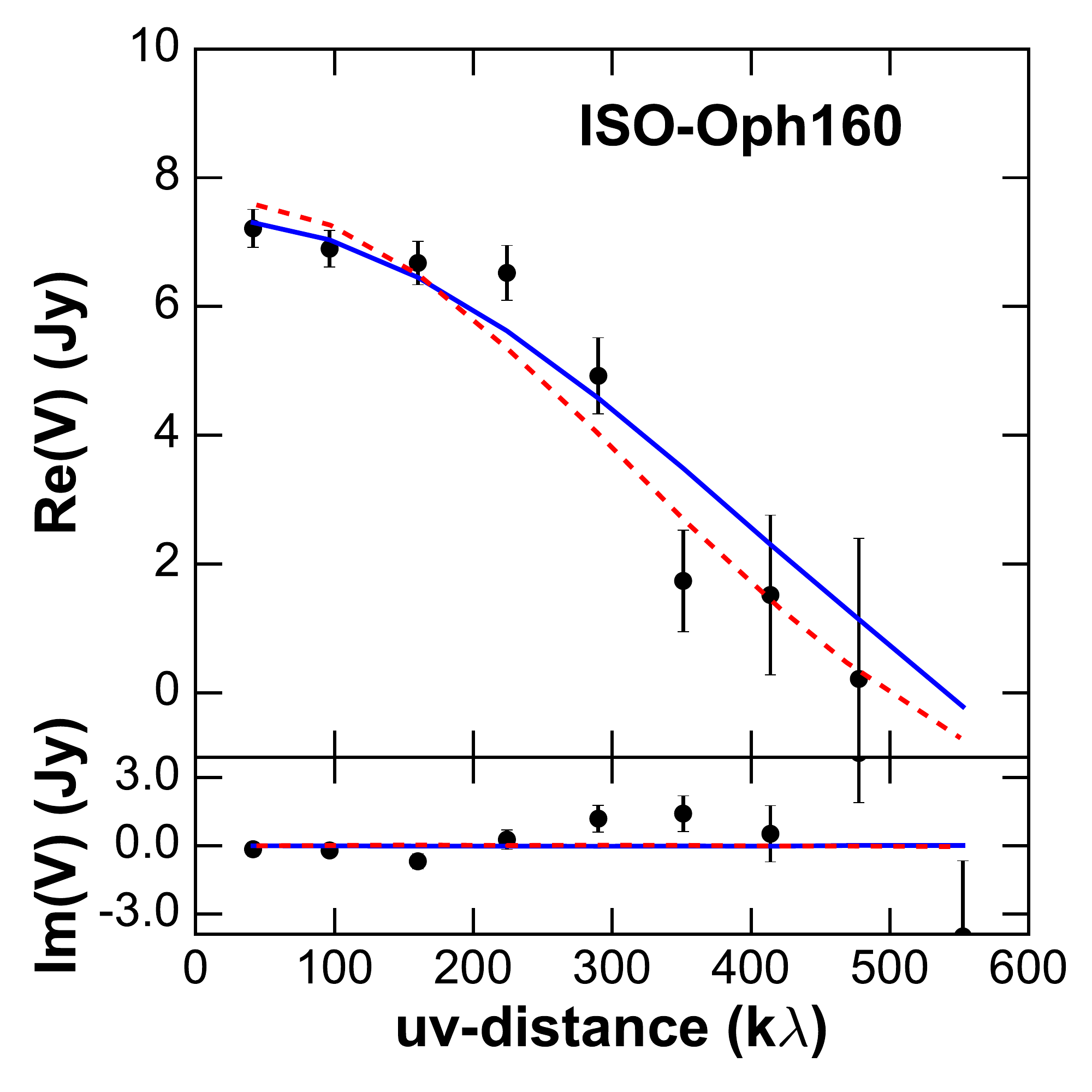}
        \includegraphics[width=7cm]{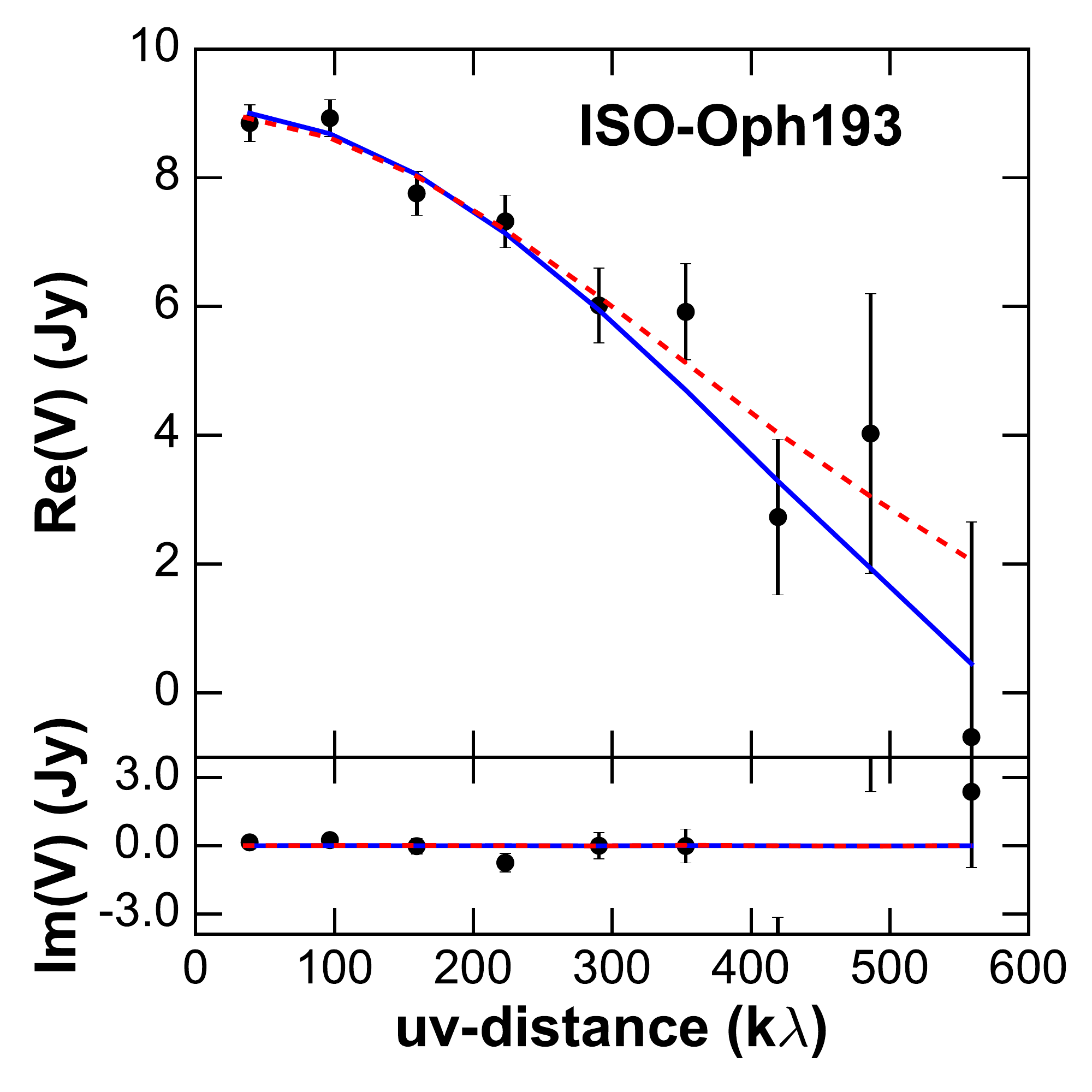}
        %\end{center}
        \caption{Real (top subpanel) and imaginary (bottom subpanel) visibilities as a function of baseline length for ISO-Oph160 (top figure) and ISO-Oph193 (bottom figure). The ALMA points were binned to increase the signal-to-noise ratio. The curves show the best-fitting models: the truncated power law $\Sigma$ model as a solid blue line and the exponentially tapered model as a red dashed line. }
   \label{f_fituvp}
\end{figure}

   \begin{figure}
        \begin{center}
                \includegraphics[width=7.8cm]{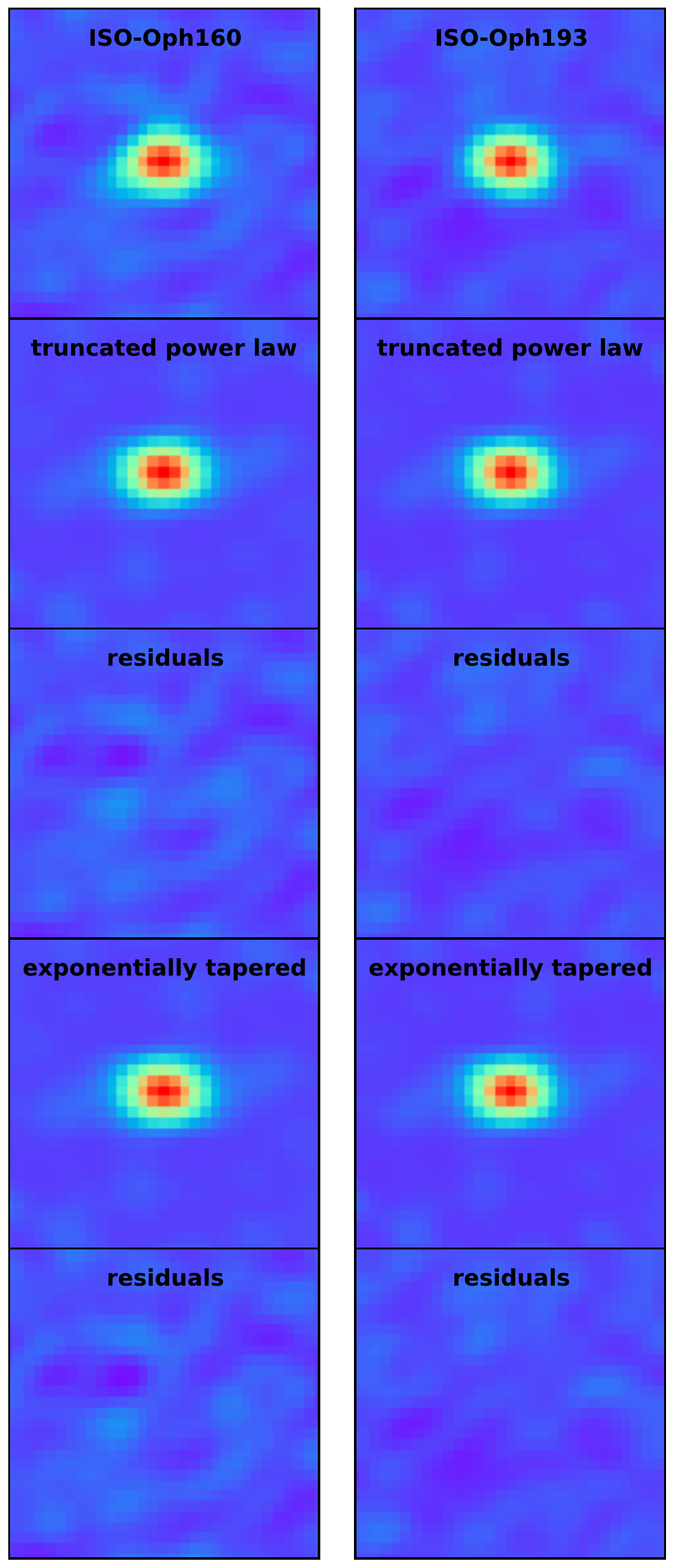}
        \end{center}
        \caption{Comparison between observations and best-fit models for ISO-Oph160 (left column) and ISO-Oph193 (right column). From top to bottom: observed, model and residual images for the best fit with $\Sigma(R)$ parametrized as a truncated power law; model and residual images for the best fit with the exponentially tapered $\Sigma(R)$. The size of each panel is $3^{\prime\prime}$ as in Fig.~\ref{f_images}}
   \label{f_fitres}
\end{figure}

Table~\ref{table_alma} gives in Column 4 the deconvolved major axis computed assuming a Gaussian source. Of the 11 detected objects, 8 are not resolved with deconvolved radii $\simless 24$ AU for the adopted distance D=122 pc.
Three objects, ISO-Oph042, ISO-Oph160, and ISO-Oph193, are marginally resolved.

%For the two brightest objects, we plot the ALMA visibilities are shown in Fig.?? for all the objects. For some disks, there is a hint that the visibilities decrease slowly at small baselines, 
%suggesting a steep decrease of the surface brightness. If we consider that, to zero order,  the brightness 
%is proportional to the product $\Sigma \times T$, given that the temperature profile is never very steep, this implies a steep decrease of the dust column density at the scales sampled with ALMA. However,
%the signal-to-noise ratio is too low, especially at the larger baselines, for any conclusion. 

ISO-Oph160 and ISO-Oph193 are detected with a high signal-to-noise ratio ($\ge 30$), hence we attempted a more careful analysis of the disk structure. Following \citetads{2016A&A...588A..53T}, we performed a fit to the visibilities using a simple two-layer disk model with two different prescriptions for the distribution of the surface density: a power law distribution with a sharp outer radius or an exponential taper, in the following we refer to these two models as truncated power law and exponentially tapered.

In the truncated power law case, $\Sigma(R)$ follows a power law of exponent $p$ out to a radius $R_{out}$ as follows:
\begin{equation}
\label{eq_pl}
\Sigma (R)= \Sigma_0 \left(\frac{R}{R_0}\right)^{-p} 
.\end{equation}
For the exponentially tapered, we adopted the parametrization of \citetads{2011A&A...529A.105G}, i.e.
\begin{equation}
\label{eq_g}
\Sigma (R)= \Sigma_0 \left(\frac{R}{R_0}\right)^{-\gamma} \> \exp\left[-\left(\frac{R}{R_c}\right)^{2-\gamma}\right]
,\end{equation}
which is a solution of the self-similar evolution of viscous accretion disks (where viscosity increases as $R^\gamma$), and is equivalent to other commonly used parametrizations \citepads[e.g.][]{2009ApJ...701..260I}.   The functional form of $\Sigma(R)$ is defined by three parameters: $R_0$ and $p$ or, equivalently, $R_c$ and $\gamma$, and $\Sigma_0= \Sigma(R_0=10AU)$, which is the normalization factor that scales with the  disk mass. In all cases we adopt an inner radius of 0.1~AU, as the results do not depend on its exact value.

%In both cases the normalization factor $\Sigma_0$, i.e., the value of $\Sigma$ at $R_0$, is such that the total disk mass is $M_{disk}$.
Once $\Sigma$ and the stellar parameters are fixed, the disk temperature and the surface brightness at 890~$\mu$m are computed using  the two-layer radiation transfer model \citepads{1997ApJ...490..368C,2001ApJ...560..957D,2001ApJ...547.1077C} with flare reduced by a factor 3 to match the typical values of the far-infrared excess  observed in BDs (Daemgen et al.~2016, submitted).
The dust opacity is calculated using Mie theory assuming the same dust composition throughout the disk, given by the following fractional abundances adapted from \citetads{1994ApJ...421..615P}: 5.4\%\ astronomical silicates, 20.6\%\ carbonaceaous material, 44\%\ water ice, and 30\%\ vacuum \citepads[see][for details]{2013A&A...558A..64T}. Furthermore, we assume a power law grain size distribution $n(a) \propto a^{-q}$ for $a_{min} \le a \le a_{max}$, where $a$ is the grain size. In order to model the fact that we expect larger grains in the disk midplane than on the surface \citepads[see][]{2014prpl.conf..339T}, we use different prescriptions for the grains in the disk surface ($a_{min} = 10$~nm, $a_{max} = 1~\mu$m, and $q = 3.5$) and for the midplane ($a_{min} = 10$~nm, $a_{max} = 1.4$~cm, and $q=3$).
The midplane dust opacity is $\kappa_{890}$ =2 cm$^2$/g.The dust-to-gas mass ratio is fixed to 10$^{-2}$.

The method  uses a Bayesian approach and implements an affine-invariant Markov Chain Monte Carlo (MCMC) ensemble sampler \citepads[see][for a full description]{2016A&A...588A..53T}. For each model we compute the 890~$\mu$m synthetic image and we sample its Fourier transform at the (u,v) plane location of the observations. We estimate the Bayesian probability using $exp(-\chi^2)$ as likelihood with flat priors.
In this scheme there are five free parameters, namely the disk mass $M_{disk}$, the two parameters that define $\Sigma$ ($R_c$, $\gamma$ for the exponentially tapered profile, $R_{out}$, $p$ for the truncated power law), the inclination, and position angle of the disk on the sky. Two additional nuisance parameters are used for the disk centre position on the sky.

\begin{table}
\caption[]{Results of the model fits for the parameter sets \{$R_c$,$\gamma$\}\ and \{$R_{out}$,$p$\}; see Sect.~\ref{s_rad} and App.~\ref{app_tau} for details. The 1$\sigma$ confidence range for each parameter is also shown.}
         \label{table_size}
%\begin{center}
%    \begin{tabular}{l c c  c c }
%        \hline
%Name &  $R_c$ & $\gamma$&$R_{out}$& $p$ \\
%          &   [AU]&   &   [AU]&   \\
%\hline

%ISO-Oph160 &  24.0&  -0.50 & 33.3&  0.0\\
%\smallskip
%ISO-Oph193 &  21.5&  -0.43 & 25.1&  -0.5\\
%\twom $^a$ & 64.1& 1.37& 154.4& 1.85\\
%\cida $^a$& 83.6& 0.85&  133.9& 1.35\\
%\ctfour $^a$& $\ge 100$& 1.0& $\ge 100$& 1.08\\ 
%\end{tabular}
%\end{center}
%$^a$) Data for the Taurus BDs is from \citetads{2014ApJ...791...20R}, model fits have been performed with the same procedure used for the \roph\ objects, see Appendix~\ref{app_tau}.
%\end{table}
%\begin{table}[]
%\caption{For publication}
\begin{center}
\begin{tabular}{lllll}
\hline
Name             &      $R_{c}$ & ${\gamma}$ & ${R_{out}}$ & ${p}$ \\
                         &  [AU]    &  &          [AU]\\
\hline
ISO-Oph160   & $     24_{-3}^{+ 2}$ & $     -1.4_{-0.2}^{+ 0.9}$ & $     22.0_{-0.4}^{+12}$ & $     -1.0_{-0.4}^{+ 1.1}$\\[10pt]
\medskip
ISO-Oph193   & $     23_{-3}^{+11}$ & $     -0.34_{-0.7}^{+ 0.8}$ & $     22_{-2}^{+16}$ & $     -0.64_{-0.02}^{+ 1.5}$\\[10pt]
\twom  & $     49_{-11}^{+48}$ & $      1.4_{-0.1}^{+ 0.5}$ & $    153_{-40}^{+100}$ & $      2.0_{-0.2}^{+ 0.3}$\\[10pt]
\cida       & $     82_{-15}^{+ 8}$ & $      0.9_{-0.1}^{+ 0.1}$ & $    136_{-45}^{+17}$ & $      1.35_{-0.13}^{+ 0.09}$\\[10pt]
\ctfour    & $>90$ & $      1.0_{-0.3}^{+ 0.2}$ & $   >90 $ & $      1.0_{-0.1}^{+ 0.6}$
\end{tabular}
\end{center}
\label{default}
\end{table}%

In Figs.~\ref{f_fitpar}, \ref{f_fituvp}, and~\ref{f_fitres} and in Table~\ref{f_fituvp}, we show the results of the fits for our two disks. The probability distributions shown in Fig.~\ref{f_fitpar} illustrate the range of the parameters \{$R_c$,$\gamma$\}\ and \{$R_{out}$,$p$\}\ derived from our models. The shaded areas correspond to $\sim 68$\%, $\sim 95$\%, and $\sim 99$\%\ confidence levels. In Fig.~\ref{f_fituvp} we show the real and imaginary part of the visibilities of the two sources and overplot two models, one for the truncated power law $\Sigma(R)$ and one for the exponentially tapered, which are selected as those with the lowest $\chi^2$. The parameters of these models are indicated with a yellow star in Fig.~\ref{f_fitpar}; \textbf{the minimum normalized $\chi^2$ range between 1.00 and 1.38}. In Fig.~\ref{f_fitres} we show the observed ALMA images for the two sources (same as Fig.~\ref{f_images}), the model images and the residual images computed from the simulated visibilities for the models indicated with a yellow star in Fig.~\ref{f_fitpar} and plotted in Fig.~\ref{f_fituvp}. In Table~\ref{table_size} we report the numerical values and uncertainties of ($R_c$,$\gamma$) and ($R_{out}$,$p$) for the sources analysed in this paper.

The modelling results suggest that the disks are compact ($R\simless$25~AU), which is also consistent with the small deconvolved sizes (Table~\ref{table_alma}). The values of $\gamma$ and $p$ are $\le 0$ and $\le -0.5$, respectively. This implies that the derived profiles of $\Sigma$ for ISO-Oph160 and ISO-Oph193 are consistent with a rather flat inner region for $R \simless 20-25$ and a very steep cutoff. They also confirm that the (sub)mm emission is optically thin at all radii.
All other disks in our sample are too faint to run this analysis and their small apparent sizes (Table~\ref{table_alma}) may either be the result of small and sharply truncated disks, as for ISO-Oph160 and ISO-Oph193, or compatible with a decline of $\Sigma$ and a larger outer radius, as our limited surface brightness sensitivity does not allow us to distinguish between these two cases \citepads[see e.g. the discussion in][]{2008Ap&SS.313..113N}.

As a cautionary note, we point out that our models have been computed with identical dust properties across the disk (as we have no direct constraint that would allow us to use a different approach). If dust grows, migrates, and eventually has a size distribution that depends on the distance from the star \citepads[e.g.][]{2010A&A...513A..79B}, then using identical dust properties throughout the disk may result in deriving a $\Sigma(R)$ profile that is shallower than the true profile \citepads[see eg. the discussion in][]{2011A&A...525A..12B,2013A&A...558A..64T}.

\section {Discussion}

\subsection{BD disk masses and potential for planet formation}

\begin{figure*}
        \begin{center}
                \includegraphics[width=13.5cm]{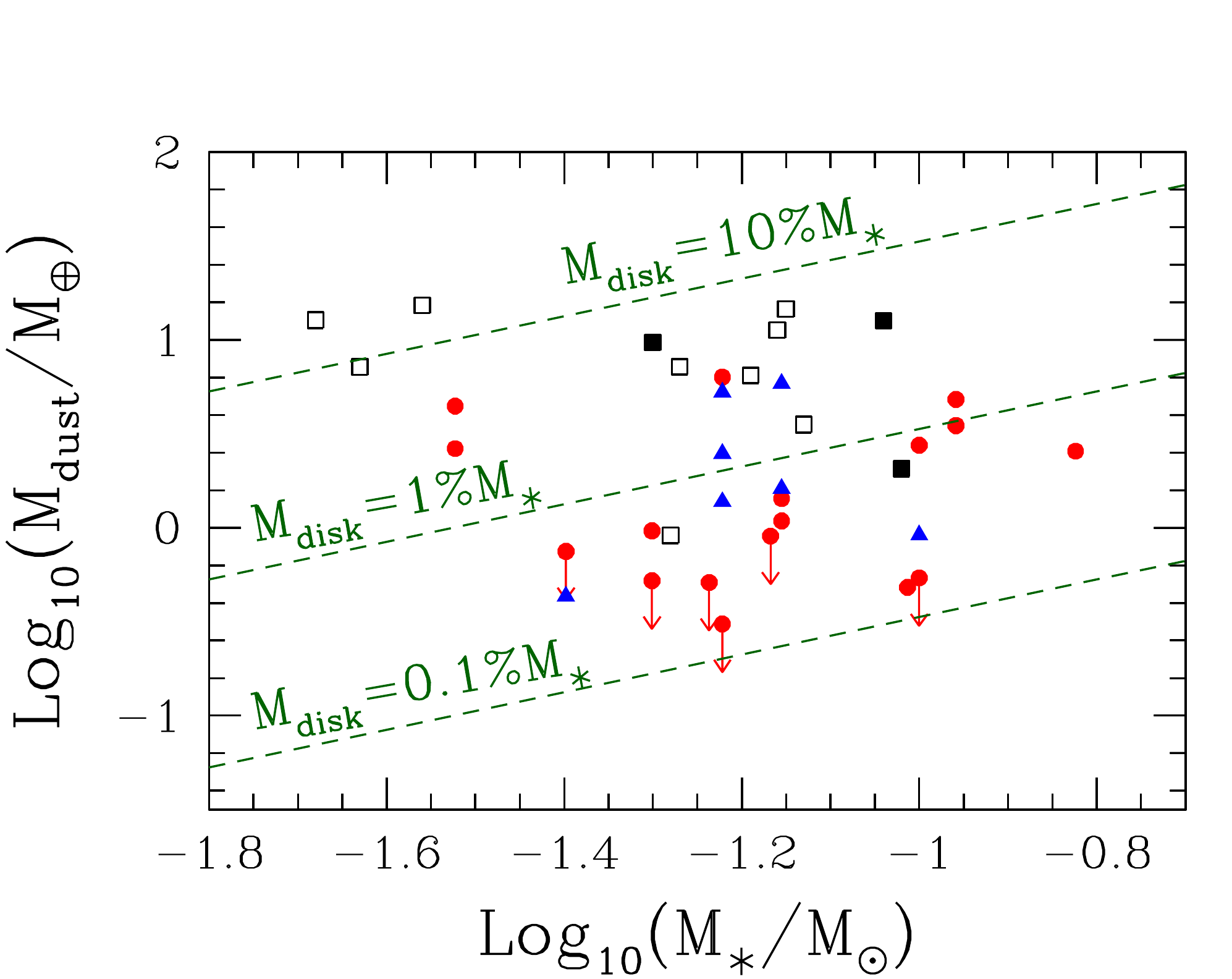}
        \end{center}
        \caption{Disk masses as function of the central object mass \Mstar.  Red dots and arrows are for the \roph\ sample discussed in this paper. The black squares report the disk masses for BDs in Taurus from the \Fmm\ detections: open, single-dish measurements from \citetads{2005ApJ...631.1134A,2014ApJ...789..155B,2013ApJ...773..168M}; filled, from ALMA measurements by \citetads{2014ApJ...791...20R}. Blue triangles show the values for the 7~BDs in Upper~Sco observed with ALMA by \citetads{2016ApJ...819..102V}.  For all objects, disk  masses have been computed as in \roph, assuming T=25(\Lstar/\Lsun)$^{0.25}$~K, \opmm=2.0 cm$^2$/g. } %$3\sigma$ upper limits are comparable to the detections.}
\label{fig_allbd}
\end{figure*}

Fig.~\ref{fig_allbd} compares the values for disk masses of BDs that we obtained in \roph\  with  disk masses of BDs in the literature  derived from (sub)mm fluxes: Upper~Sco \citepads{2016ApJ...819..102V} and in Taurus (see figure caption for references). For consistency, we recomputed all disk masses using Eq.~\ref{eq_1} with the appropriate $T_A$ for each object and \opmm =2 \cmg.  The values of \Mdust\ in the \roph\ and Upper~Sco objects cover a similar range, while several objects in Taurus have \Mdust $\sim$ 10\Mearth.
 Most Taurus BDs have been observed with single-dish telescopes and, in principle, the large fluxes could be the result of diffuse emission contamination. However, when both single dish and interferometric fluxes are available, they do not differ
significantly \citepads[see e.g.][]{2007ApJ...671.1800A,2012ApJ...761L..20R,2014ApJ...791...20R}. 

The largest ratios \Mdisk/\Mstar\  are similar in the three regions, i.e. between 0.01 and 0.1. However, contrary to Taurus and U Sco,  \roph\  has  a large number of BD disks with \Mdisk/\Mstar $\simless$ 0.1. This may appear as a surprising result, given that \roph\ should be the youngest of these regions. This difference is likely  a selection effect, which is similar to what was already noted by \citetads{2006A&A...452..245N} for the mass accretion rate distribution. While our \roph\ sample is unbiased with no pre-selection that could favour more massive disks, this is not the case of the Upper~Sco sample \citepads{2016ApJ...819..102V}. %{\bf see Aleks?}
%In the case of Taurus, there are in the literature at least $\sim$ 40 (??) upper limits, similar in value to the actual detections, that could completely change the statistics.

Overall, the low dust masses shown in Fig.~\ref{fig_allbd}  agree with the  estimates of disk masses derived from Herschel fluxes for a number of BDs in various star-forming regions \citepads{2012ApJ...755...67H,2013A&A...559A.126A,2015A&A...573A..63L,2015A&A...582A..22L}. With the bulk of the disk population containing about one Earth mass of solids,    the ability of these disks to form planets is obviously questionable. The study by \citetads{2007MNRAS.381.1597P} suggests that disks containing a fraction of a Jupiter mass of material (including gas and dust) and surface density profiles that are shallower than $\Sigma\propto R^{-1}$ are very unlikely to form planets at all. If we assume a gas-to-dust ratio by mass of $\sim 100$, only the most massive disks in Fig.~\ref{fig_allbd} (those with M$_{dust}\sim 10$M$_\oplus$) will be massive enough to form planets. In addition, if the $\Sigma(R)$ profiles derived
in Sect.~\ref{s_rad} are found to be common in BD disks, the prospects of planet formation in these systems is really grim. Some optimism may be still possible for some of the most massive BD disks in Taurus \citepads[see][and Sect.~\ref{s_sdisc}]{2014ApJ...791...20R}, such disks may explain the detection through microlensing of Earth-mass planets around BDs \citepads[e.g.]{2015ApJ...812...47U}. More massive, supra-Jupiter mass companions \citepads[e.g.]{2004A&A...425L..29C,2007ApJ...666L.113J,2010ApJ...714L..84T,2013ApJ...778...38H} cannot be really considered to be formed in protoplanetary disks with the properties that we observe, and are most likely the result of a binary system formation in the substellar regime.

%{\bf Comments on Lodato papers? Little mass and flat Sigma: disaster!}. 

\subsection{Comparison between BDs and TTS disk masses}

\begin{figure*}
        \begin{center}
        \includegraphics[width=13.5cm]{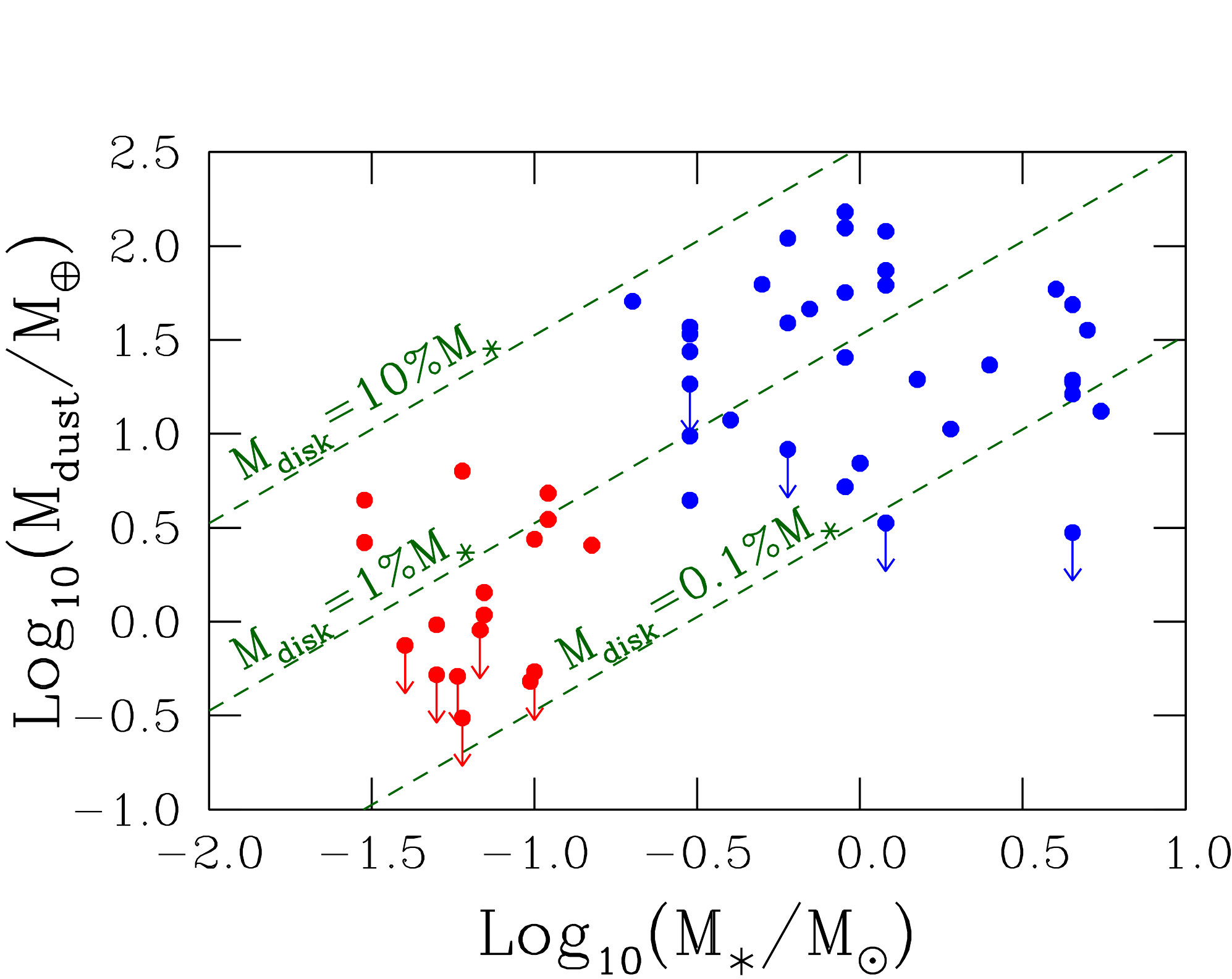}
        \end{center}
        \caption{Dust mass vs. stellar mass  for \roph\ BDs (red) and TTS (blue). %The black open symbols are from 1.3mm fluxes. 
    The dashed green lines show the dust mass in a disk with 10\%, 1\%, 0.1\% the stellar mass, computed assuming a gas-to-dust mass ratio of 100. }
\label{fig_ophall}
\end{figure*}

%The situation is different for Taurus, for which we find in the literature mm detections of about  20 BDs, and about 40 upper limits of comparable magnitude. We have computed the disk mass from the observed fluxes as for the \roph\ objects. In Taurus,

%The  mass distribution of disks around BDs when compared to that of disks around TTS in the same star forming region  can give us clues on the process of BD formation, as  discussed in the introduction. 
In Fig.~\ref{fig_ophall} we show  the disk mass for the \roph\ BDs together with the disk mass for the TTS in the same region as a function of the mass of the central object. To compute the disk masses for the TTS sources, we used the 850 $\mu$m fluxes from \citetads{2007ApJ...671.1800A}, calculated the appropriate $T_A$ for each object, and applied  Eq.~\ref{eq_1} with the same value of \opmm\ as for the BDs. This ensures that the same method has been used to compute the masses across the whole range, the choice of using the same dust opacity for all sources 
is justified by the lack of evidence for a systematic difference in the dust properties in disks as a function of stellar mass in \roph\ and in other star-forming regions \citepads{2010A&A...521A..66R,2012ApJ...761L..20R,2014prpl.conf..339T}.
%This is justified by the fact that the mm opacity dependence on wavelength is similar (and quite flat) in objects of all mass (Ricci et al. ??). 

Fig.~\ref{fig_ophall} shows that the dust masses in disks around BDs ad TTS roughly scale as the mass of the central object. If we correct for a gas-to-dust ratio of 100 by mass, then the disk contains approximately 1\%\ of the mass of the system with a large scatter at any given mass of about one order of magnitude. A number of caveats are in order. Most importantly,  the stellar parameters of the \roph\ TTS are very uncertain, as no systematic study of TTS comparable to that carried out by \citetads{2015A&A...579A..66M} for the BDs is available. We estimated \Lstar\ and \Mstar\ from the spectral types of \citetads{2010ApJS..188...75M}, assuming that all the stars lie on a 1~Myr isocrone as determined using the theoretical evolutionary tracks of \citetads{1998A&A...337..403B}. Although this does not affect the classification as TTS, which is based on the spectral type, it affects the exact location on the diagrams and the estimate of the disk-to-star mass ratio. Secondly, the sample of BDs and TTS that have good sub-mm measurements is largely incomplete. The completeness of the BDs sample in this study has been addressed in Sect.~\ref{s_sample}; the TTS sample limitations is described in \citetads{2007ApJ...671.1800A}. In this paper we  proceed with the analysis under the assumption that the samples are representative of the properties of the full populations, but this is an assumption that needs to be put on firmer ground with future optical and near-infrared spectroscopic studies of the central stars and ALMA surveys of the full populations.

\begin{figure}
        \begin{center}
        \includegraphics[width=8cm]{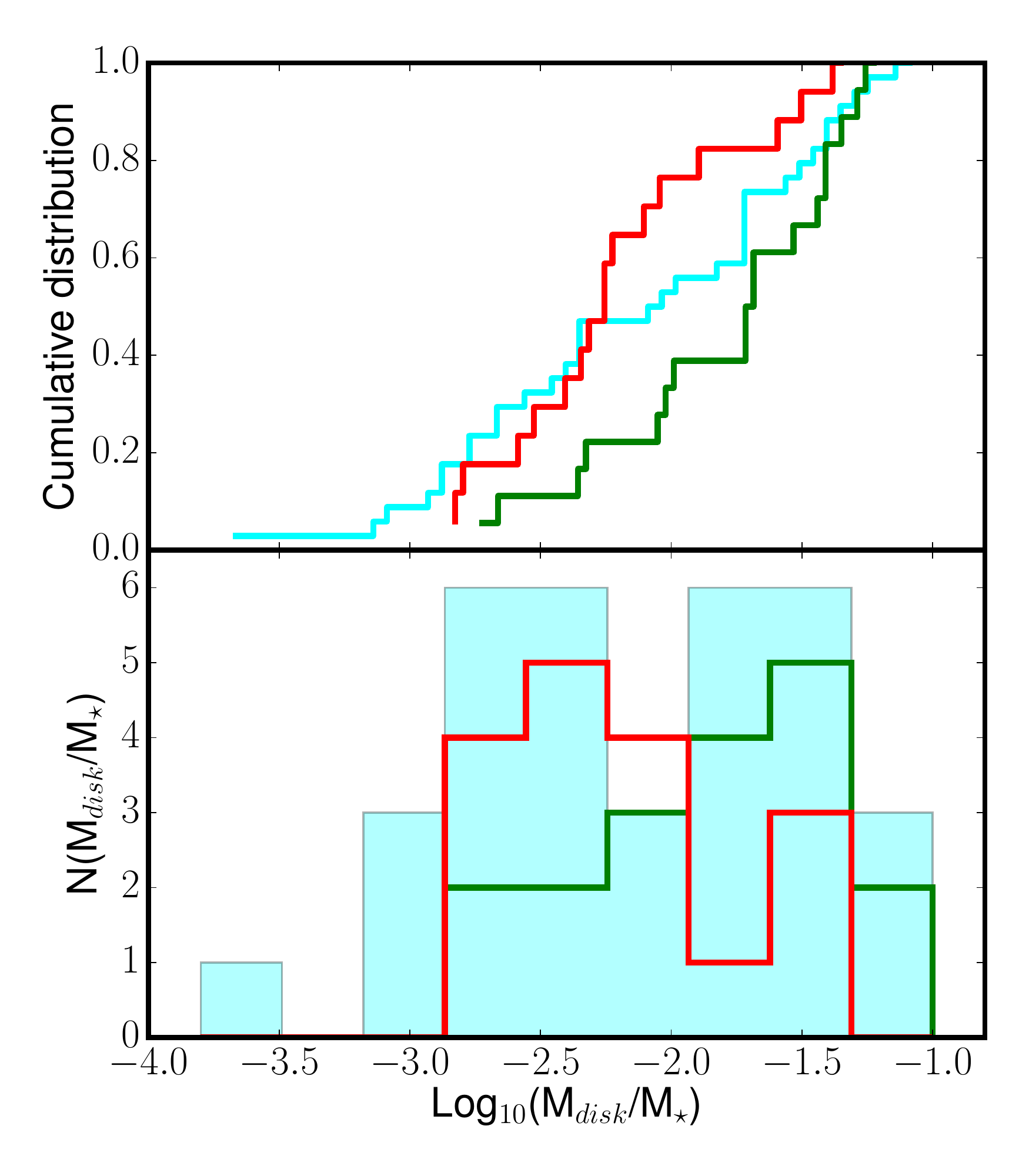}
        \end{center}
        \caption{Cumulative distributions (top) and histogram of the M$_{disk}$/M$_\star$ values for \roph\ BDs (red), the full TTS sample (cyan), and the subset with $0.2\le\rm M_\star/M_\odot\le 1.0$ (green). 
    %Although the full TTS distribution appears broader than the BDs distribution, the two samples are consistent with being drawn from the same parent distribution, according to the Anderson-Darling test.
    }
\label{fig_ophcum}
\end{figure}

In Fig.~\ref{fig_ophall} there is an apparent paucity of BD disks in the region 1-10\%\ of the mass of the central object compared with TTs. To quantitatively estimate whether there is a real discrepancy between the two samples, we  analysed the cumulative distributions of the ratio M$_{disk}/$M$_\star$ for BDs and TTS. In Fig.~\ref{fig_ophcum} we show (in the bottom panel) the  distributions of the ratio M$_{disk}$/M$_\star$ for the full \citetads{2007ApJ...671.1800A} sample (cyan filled histogram), a subset of that sample with $0.2\le\rm M_\star/M_\odot\le 1.0$ (green histogram), and for our BDs sample (red histogram). The top panel of Fig.~\ref{fig_ophcum} shows the cumulative distributions for the same samples. The TTS sample contains objects with central mass in the range from $\sim$0.2~M$_\odot$ to $\sim$5~M$_\odot$.
 We performed the Anderson-Darling test\footnote{We used the {\tt scipy.stats} python implementation of the test, which follows \citetads{Scholz:1987:SAD}}
between these two samples and find $\sim$1.6\%\ probability that the two are drawn from the same parent distribution to check whether the sample of disk masses for the disks of the BDs is drawn from
a different distribution than the solar-mass sample (for the purpose of this test we treated upper limits as measurements). The major caveat to bear in mind is the fact that the samples may not be fully representative of the full populations and that the methods for deriving the disk and stellar masses in the two samples are not fully identical. Future detailed characterization of, in particular, the TTS population (both the disk and photospheric properties) will allow us to provide a definitive answer as to whether the possible difference we see here is real.

\subsection{Comparison with Taurus BD disks structure}
\label{s_sdisc}

The two clearly resolved \roph\ disks  both  show  a sharp dust disk edge at $\sim$ 20-30 AU with a rather flat surface density profile inside it (see Sect~\ref{s_rad}, and Table~\ref{s_sdisc}). While this second aspect may be due to our assumption that the dust properties do not change with radius, the sharp outer edge of these two disks seems to be a robust result.
This finding is in  contrast with the results on BD disks sizes reported by \citetads{2014ApJ...791...20R} for disks in Taurus. To provide a proper  comparison, we re-analysed the three~BD disks in Taurus using the method outlined in Sect.~\ref{s_rad}. In Table~\ref{table_size} we show the comparison between the values of \{$R_c$,$\gamma$\}\ and \{$R_{out}$,$p$\}, which are derived by the two families of models, for the \roph\ and Taurus BDs (the figures with the model results for Taurus are reported in App.~\ref{app_tau}). We confirm and extend the results of the previous study. The truncated power law parameters from our reanalysis are consistent with previous results and, in addition, the exponentially tapered models are also consistent with a decreasing $\Sigma(R)$ and a smooth outer disk edge. This seems to suggest the possibility of an important difference between the two most massive BD disks in \roph\ and those in Taurus. The Taurus disks have a structure very similar to those of more massive TTS, while the \roph\ BDs have smaller disks with a sharper outer edge. The dust mass in the disks does not seem to be very different, $\sim 4.8--6.3$ \Mearth\ in Oph and $\sim 2, \sim 10,$ and  $\sim 15$ \Mearth\ for \ctfour, \twom\ and \cida, respectively.

Both the formation of BDs as ejected embryos \citepads{2001AJ....122..432R} and dynamical interactions in a dense star-forming region would predict that disks may be truncated to small radii at some point of the early evolution of the system \citepads{2009MNRAS.392..590B,2012MNRAS.419.3115B}. In all cases it is expected that the gaseous component of the disk viscously re-expands on timescales shorter than $\sim$1~Myr or a new outer disk could be accreted from the cloud \citepads{2012MNRAS.419.3115B,2014A&A...566L...3S}. 
At face value, our numbers are consistent with the results of \citetads{2012MNRAS.419.3115B} models; these models show that the distribution of disk sizes in young dense clusters may be rather small ($\simless$10~AU for 50\% of the disks), but with a significant tail that extends to R$\simgreat$100~AU. The low density and/or slightly older age of Taurus may naturally explain the larger radii for \twom, \cida, and \ctfour. 

According to \citetads{2009MNRAS.392..413S}, BDs formed in the disk of a solar-mass star have disks with  masses of $\simless$few \MJ\ and radii of 20-50 AU with a small fraction of larger disks. There is no correlation between the BD disk properties and those of the parental star. These disks are likely stripped by the following ejection, and the distribution of properties
of the surviving disks is not clearly predicted by models. It is possible that both the \roph\ and the Taurus BDs can be explained by ejection scenarios. Lacking large observational samples and more detailed theoretical models that describe the properties and evolution of disks after ejection, we cannot really constraint in detail the BDs formation scenario.
%\LEt{This was ambiguously worded. Please check for meaning (For example, I wasn't sure what you meant by BD/disk formation. It needs some kind of qualifier, such as "extent of"\ or "possibility of".}

%{\bf Add or not comments on: 
%sizes from mm grains, I suspect that a truncated disk may re-expand the gas more efficiently than the grains responsible for the mm flux. This would suggest that dynamical truncation in Oph (rather than ejection everywhere and re-expansion in Taurus) is the likely cause of what we see. Note our estimate of the disk size in Oph 102 was relatively large (>~ 40AU) from gas. Need to check what we have from dust. 
%\item Samples are still very small, we cannot really compare statistically with Bate12
%} 

Dynamical truncation of the disk radius on timescales longer than  the viscous lifetime may be caused by the presence of a companion, which may truncate the circumstellar disk at a radius of $\sim$0.3--0.5 of the binary separation and effectively prevent its viscous expansion \citepads{1994ApJ...421..651A}. There are no reported companions for ISO-Oph160 and ISO-Oph193, but none of the current infrared surveys would be sensitive to close ($\sim 0.\!^{\prime\prime}5-1.\!^{\prime\prime}0$), planetary-mass companions. A dedicated measurement would be in order before ruling out this possibility. As a plausibility argument, we note that in our survey we do not find any large and massive disks in \roph\ that resemble those found in Taurus, and we expect that our unbiased search would have not missed these large, smooth disks. 

A sharp edge in the dusty disk, as observed at millimetre wavelengths, can also be the result of a very different process, namely grain evolution. As grains grow, fragment and drift, the large grains, that are responsible for the observed mm emission, decouple from the gas, and concentrate more towards the star.
This effect, which is predicted by global disk evolution models, is also measured in large protoplanetary disks around TTS and Herbig Ae/Be stars \citepads[e.g.][]{2010A&A...513A..79B,2012ApJ...760L..17P,2015arXiv151205679T,2014prpl.conf..339T}.
\citetads{2014ApJ...780..153B} have shown that as a result of this process,
the large grains surface density, as probed by millimetre observations, shows a sharp decrease  at a radius significantly smaller than that of  the gaseous disk. The edge of the dusty disk, however, also tends to get smoother with time (on viscous timescales).  Similar results are also found when  photoevaporation is taken into account \citepads{2015ApJ...804...29G}. Although in this case the difference in size between the gaseous and dusty disk is reduced, the mm-size grains have a sharper outer cutoff than the gas. 
There is plenty of evidence of grain evolution in BD disks \citepads{2004A&A...426L..53A,2012ApJ...761L..20R,2014ApJ...791...20R}. In particular, the ISO-Oph102, \twom, \cida, and \ctfour, observed with ALMA at two wavelengths by \citetads{2012ApJ...761L..20R,2014ApJ...791...20R} have mm spectral indexes $\sim 2-2.4$, and an exponent of the mm dust opacity law $\beta \sim 0.2$ that is typical of very large grains. %ISO-Oph102 has a very small dusty disk and is not resolved in our data (R<12 AU). 
Unfortunately, at present we have no information for the sample of the \roph\ BDs at other wavelengths, and we do not know if ISO-Oph160 and ISO-Oph193, in particular, have evidence of evolved, large grains. The only \roph\ object for which this information is available  is ISO-Oph102, which shows evidence of evolved grains \citetads{2012ApJ...761L..20R} and  a very compact
continuum  (deconvolved half-major axis $<12$ AU) and CO(J=3-2) emission ($\sim 15$AU), which is consistent with the results for the two resolved objects.
The Taurus BDs discussed in this paper all have evolved their grains;  their dusty disks  are large, have smooth edges, and are surrounded by extended gaseous disks, as TTS. If dust evolution occurs in both \roph\ and Taurus BD disks, %is responsible for the sharp edges of the two \roph\ BDs, 
then the evolutionary stage or path of the two groups must be different, possibly owing to very different ages, initial disk structures, and/or environmental effects.
However, this remains to be proved. On one side, the number of objects  studied so far is very small and needs to be significantly increased. On the other, the difference between the two groups is significant at a 2.5$\sigma$ level only, and ALMA observations with better resolution and sensitivity are in order.

\section {Conclusions and summary}

Our millimetre observations of disks in \roph\ BDs provides the first large, homogeneous, and well-characterized sample in a young star-forming region. The sample is unbiased, as it contained all the spectroscopically confirmed Class II BDs at the time of the proposal, with no selection that could favour more massive disks. We detected 11/17 ($\sim 65$\%) objects;  the disk masses are  $\sim 0.5- 6.3$ \Mearth.  For the remaining 6, we set 98\%\ confidence level upper limits at $\sim 0.3-0.9$ \Mearth. 
There may be evidence for BDs to have typical values of \Mdisk/\Mstar\ that are smaller than in TTS, which possibly supports the idea of a different formation path, but the biases and limited accuracy of the data for the TTS sample do not allow us to reach a firm conclusion.
In all cases the emission from the BD disks in \roph\ appears to be compact with deconvolved major axis $\simless 20$ AU. In most cases, however, we are limited by sensitivity in our ability to detect a possible emission from a declining dust surface density in the outer disk. Two objects are resolved, and we performed a detailed analysis to derive their surface density profile. Both objects have a flat surface density and a sharp cutoff at about  $\simless 25$ AU.

In comparing with model expectations, we conclude that it should be very difficult to effectively form planets in the disks around most BDs. If the disk masses that we measure are typical, then the supra-Jupiter mass companions detected around some BDs are most likely the result of binary formation rather than bona fide planets formed in protoplanetary disks. 

The difference between the \roph\ BDs studied here with respect to the Taurus BDs studied by \citetads{2014ApJ...791...20R}, which have only slightly more massive disks (by factor $< 3$), is striking. The Taurus BD disks are much larger and have a rather smooth edge (see Table~\ref{table_size}) that is very similar to TTS disks. It is possible that the \roph\ environment is responsible, either by giving origin to disks with very different properties or because close encounters between disk systems are much more frequent than in the more diffuse region Taurus. The possibilities that the Taurus BDs had the time to viscously spread-out their disks, or that the two \roph\ BDs are close binaries cannot be ruled out with the current observational constraints.

Dust evolution (i.e. grain growth) has been confirmed both in the Taurus BDs and in ISO-Oph102, which is one of the BDs in our sample. The combination of growth, fragmentation, and drift can create dust disks that are much smaller than the gaseous disks  with relatively sharp edges. However, it is difficult to account for the difference with the Taurus BDs via dust evolution alone.

The results we found with this continuum survey are tantalizing, but clearly need to be put on firmer ground with future ALMA observations. On the one hand, it would be important to compare, with adequate signal to noise and angular resolution, the gaseous and dusty disks for the BDs with and without evidence for disk truncation discussed in Sect.~\ref{s_rad}. On the other hand, it is also necessary to put on firm grounds and for other star-forming regions the possible difference in the M$_{disk}/$M$_\star$ distributions of TTS and BDs discussed in Sect.~\ref{s_sdisc}. ALMA is now starting to provide extensive surveys of the protoplanetary disks populations around TTS in various star-forming regions (Ansdel et al. 2016, submitted; Barenfeld et al. 2016, submitted; Pascucci et al. 2016, submitted), and these offer a prime opportunity for a full extension to the substellar domain.

%The low disk masses of the BDs make it very difficult to account for the growth of grains to mm  sizes and the formation of larger bodies, if this process occurs in the disks as we observe now (Pinilla et al.??). However,  in the only \roph\ BD disk observed so far at more than one mm wavelenght (ISO-OPH102), we have evidence that grains are  very big (Ricci et al. 2013).  At the same time, the ratio of the disk mass to the mass of the central object is never larger than about few\%, contrary to what is found for TTS in the same region, where the ratio is about 10\% for a significant number of cases (???). A way to reconcile these various aspects is if BDs form not from the collapse of an individual, low mass core but as a side product of the formation of more massive objects, either in a multiple system or inside the disk of a TTS. The growth to large grains should then occur before the ejection of the BD (and its disk) from its birthplace. Of course, this is for the moment a weak evidence, which, however, we think should be explored further.  Measurements of the mm spectral index (and of the opacity dependence on wavelength) could be easily obtained with Alma, as well as spatially resolved data for all the BD disks, which could allow to reduce the uncertainty on  the disk mass. In addition, a careful study of the TTS sample should not be further delayed!

\begin{acknowledgements}
We thank Til Birnstiel for insightful discussions on the effect of grain growth on the measured outer edges of disks and Carlo Manara for the relentless efforts to overcome the difficulties in estimating correctly the photospheric parameters of young BDs.
This paper makes use of the following ALMA data: ADS/JAO.ALMA\#2012.1.00037.S and ADS/JAO.ALMA\#2011.0.00259.S. ALMA is a partnership of ESO (representing its member states), NSF (USA) and NINS (Japan), together with NRC (Canada), NSC and ASIAA (Taiwan), and KASI (Republic of Korea), in cooperation with the Republic of Chile. The Joint ALMA Observatory is operated by ESO, AUI/NRAO,kk and NAOJ.
AN  acknowledges funding from Science Foundation Ireland (Grant 13/ERC/I2907).
This work was partly supported by the Gothenburg Centre for Advanced Studies in Science and Technology  as part of the GoCAS program {\it Origins of Habitable Planets} and by the Italian Ministero dell’Istruzione, Universit\`a e
Ricerca through the grant Progetti Premiali 2012-iALMA (CUP C52I13000140001).
\end{acknowledgements}

\bibliography{rhoph_bds}
\bibliographystyle{aa}

\appendix
\section{Temperature}
\label{app_t}
\begin{figure}
        \begin{center}
        \includegraphics[width=9cm]{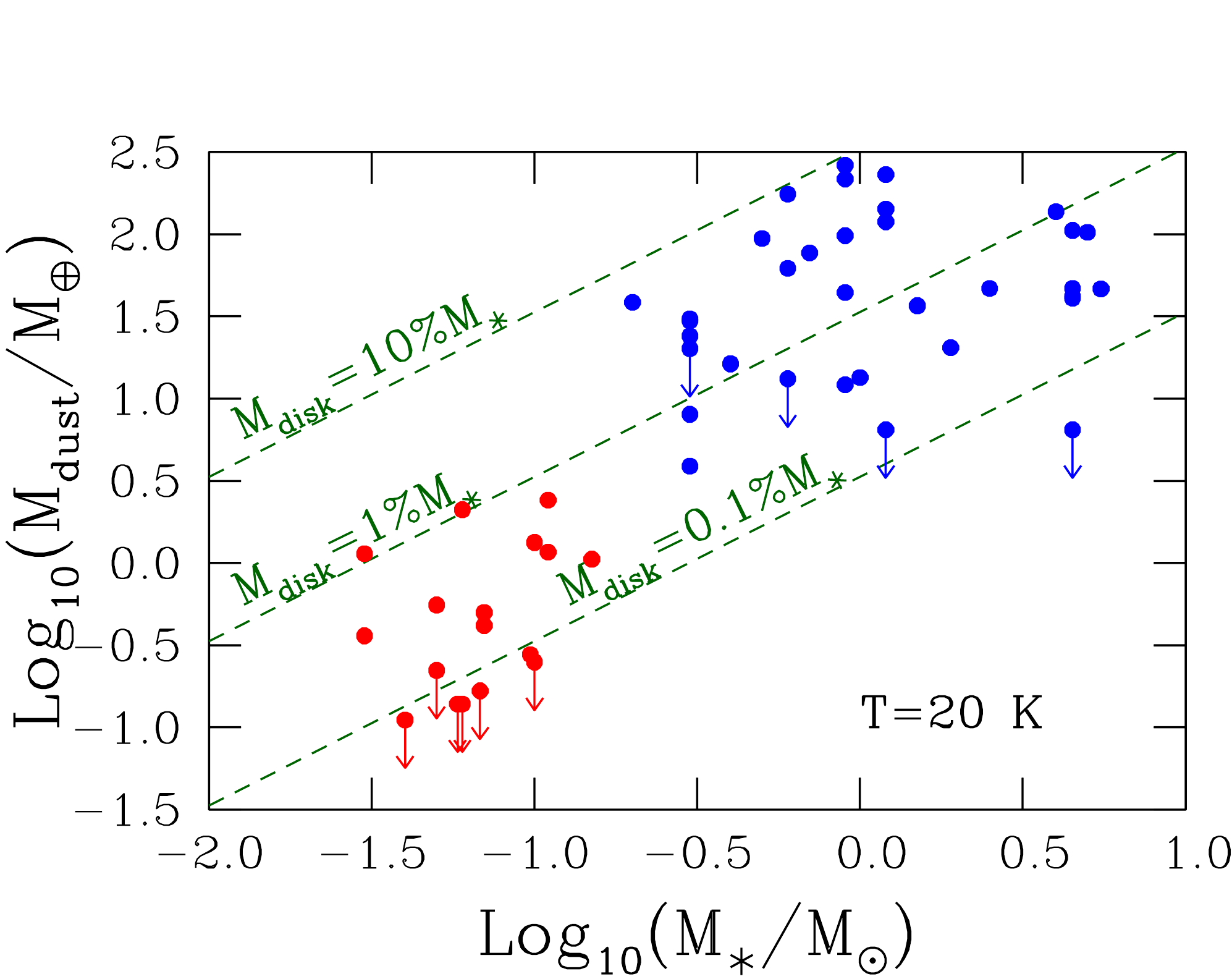}
    \includegraphics[width=9cm]{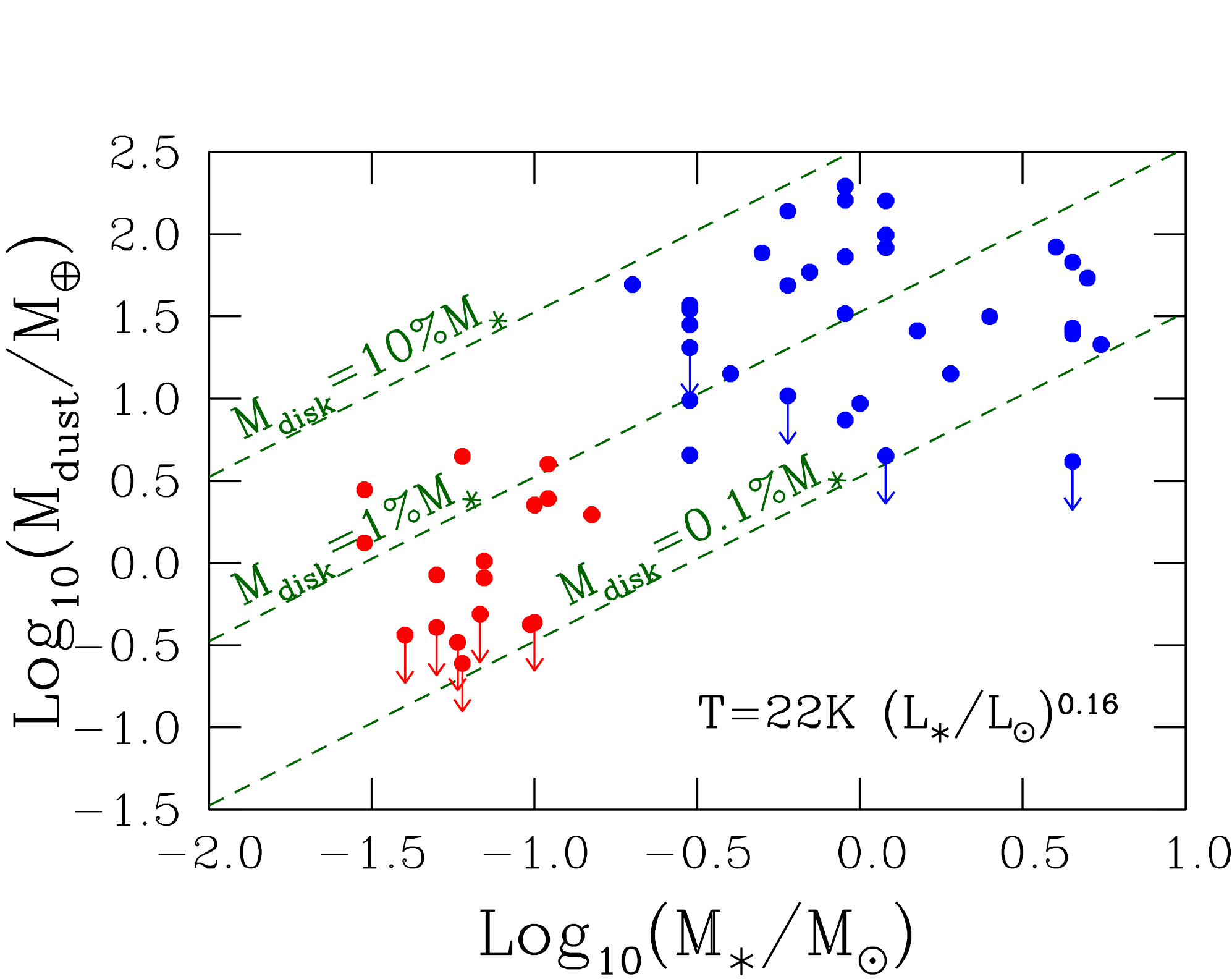}
    \includegraphics[width=9cm]{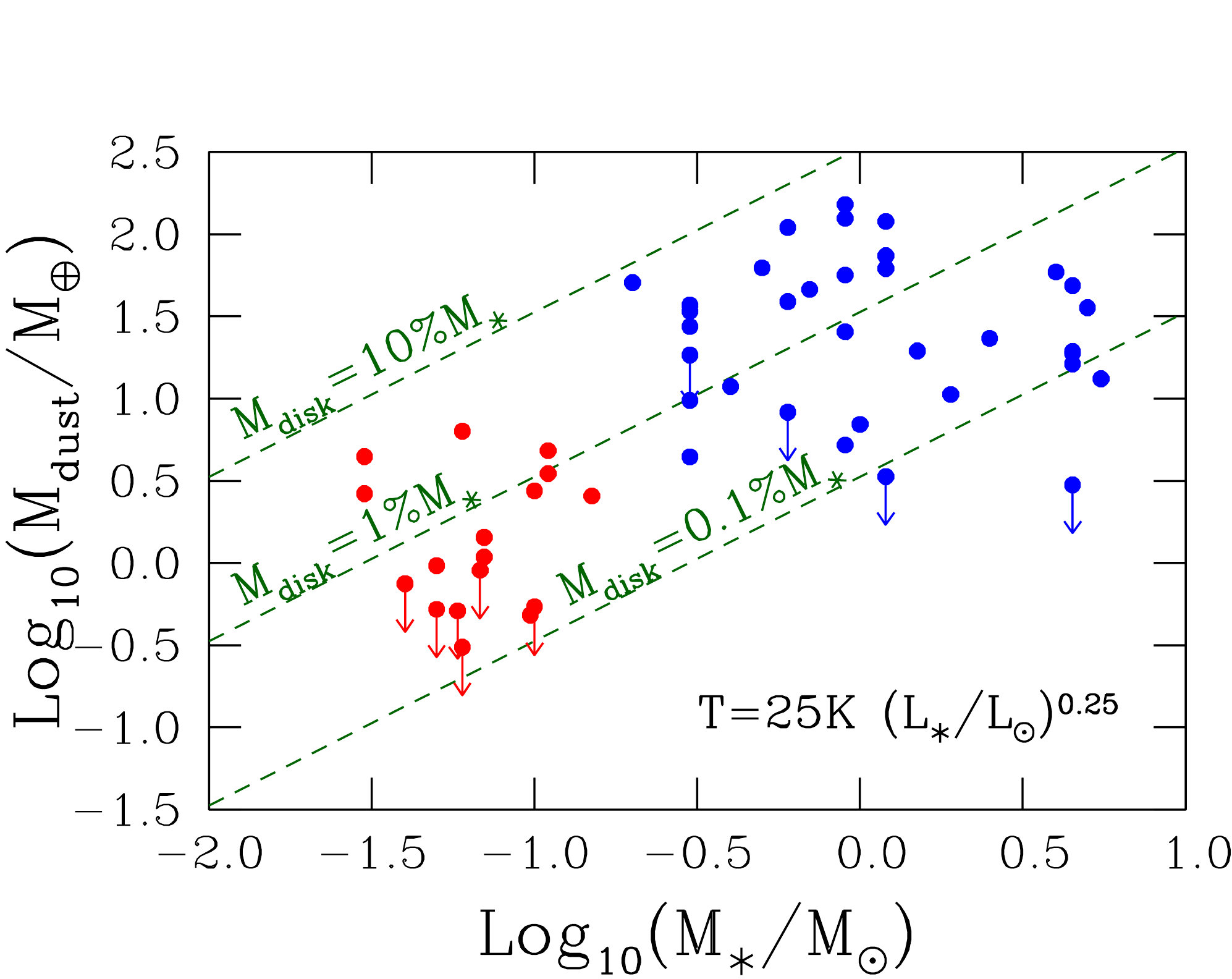}
        \end{center}
        \caption{Dust mass vs stellar mass  for \roph\ BDs (red) and TTS (blue). The top panel shows the results obtained by computing dust masses from eq.(1) with T=20K for all objects. The mid panel when adopting the scaling law of $T$ with \Lstar\ suggested by \citetads{2016ApJ...819..102V}   . The bottom panel adopting the scaling law of Andrews et al. (2013), as in Sec.4.1. %The black open symbols are from 1.3mm fluxes. 
    The dashed green lines show the dust mass in a disk with 10\%, 1\%, 0.1\% the stellar mass, computed assuming a gas-to-dust mass ratio of 100. }
\label{fig_ophall_T}
\end{figure}

The \citetads{2013ApJ...771..129A}  scaling law was derived from a grid of disk radiation transfer models over a range \Lstar $\sim 0.1-100$\Lsun. The choice of the value of $T$ appropriate
to recover the dust mass from the observed (sub)mm flux for very low-mass TTs and BDs was recently discussed by  Daemgen et al. (2016, submitted), who show that, in addition to \Lstar, the value depends strongly on the degree of disk flaring, and, to a lesser degree, on the disk mass itself and other disk parameters. Using the ratio 
of the far-infrared flux to the J-band flux, which is a good proxy for the stellar luminosity, these authors find that most objects  with Herschel far-infrared measurements have ratios that are much lower than the values predicted by  fully flared disk models. This is also true also for the BDs in our sample with measured Herschel fluxes. Based on the analysis of Daemgen et al. (2016, submitted), we conclude that the most appropriate value of the disk averaged temperature for the estimate of the disk mass from the millimetre flux is likely to be a few degrees lower than $T_A$. 
The error on the disk mass computed with temperature $T_A$, however, is of order $\pm 20$\%, with a systematic trend of underestimating the dust mass in BDs and overestimating it in low-luminosity TTS .

Very recently, \citetads{2016ApJ...819..102V} have proposed a different scaling law of $T$ with \Lstar ($T_{vdP}=22$ (\Lstar/\Lsun)$^{0.16}$), using a sample of eight BDs and very low-mass TTS in U Sco with measured ALMA fluxes. They derive the best disk mass from fitting the SEDs at all wavelengths and compare it with the results obtained from the (sub)mm flux only, using $T_A$ and their improved prescription. They find that, on average, the masses derived using $T_A$ are a factor $\sim 3$ larger than the result of the SED fits; the discrepancy is significantly reduced when $T_{vdP}$ is used. 
In practice, for our sample the difference in \Mdust\ derived using the two temperature prescriptions is not large; this difference is typically within a factor 1.5 for the majority of objects  and
$\sim 2$ for the two lowest luminosity objects, ISO-Oph033 and GY92-320. 
Given the large number of free parameters and uncertainties, and for an easier comparison with literature results for TTS, in this paper we have used $T_A$ to compute dust masses with the caveat that  the dust mass in the lowest luminosity objects may be overestimated by a factor of  up to 2.

Even if the uncertainty on individual measurements is not very large, the choice of $T$ in Eq.(1) may introduce a systematic trend  in the  disk mass--stellar mass relation, discussed in Sect. 5.2, because  in star-forming regions \Mstar\ is roughly correlated with \Lstar. In this context, the choice of $T_A$ to compute dust masses is the most conservative, i.e. this choice minimizes any existing trend of \Mdust\ with \Mstar. Fig.~\ref{fig_ophall_T} shows the results for a constant value of $T$=20K, and using $T_{vdp}$. For $T=const.$, \Mdust\ is $\propto$\Fmm. One can see that in both cases the difference between the BDs and the TTS is larger than for $T=T_A$.

\section{Model fits to the structure of Taurus BD disks}
\label{app_tau}

\begin{figure}
        %\begin{center}
  \includegraphics[width=4.4cm]{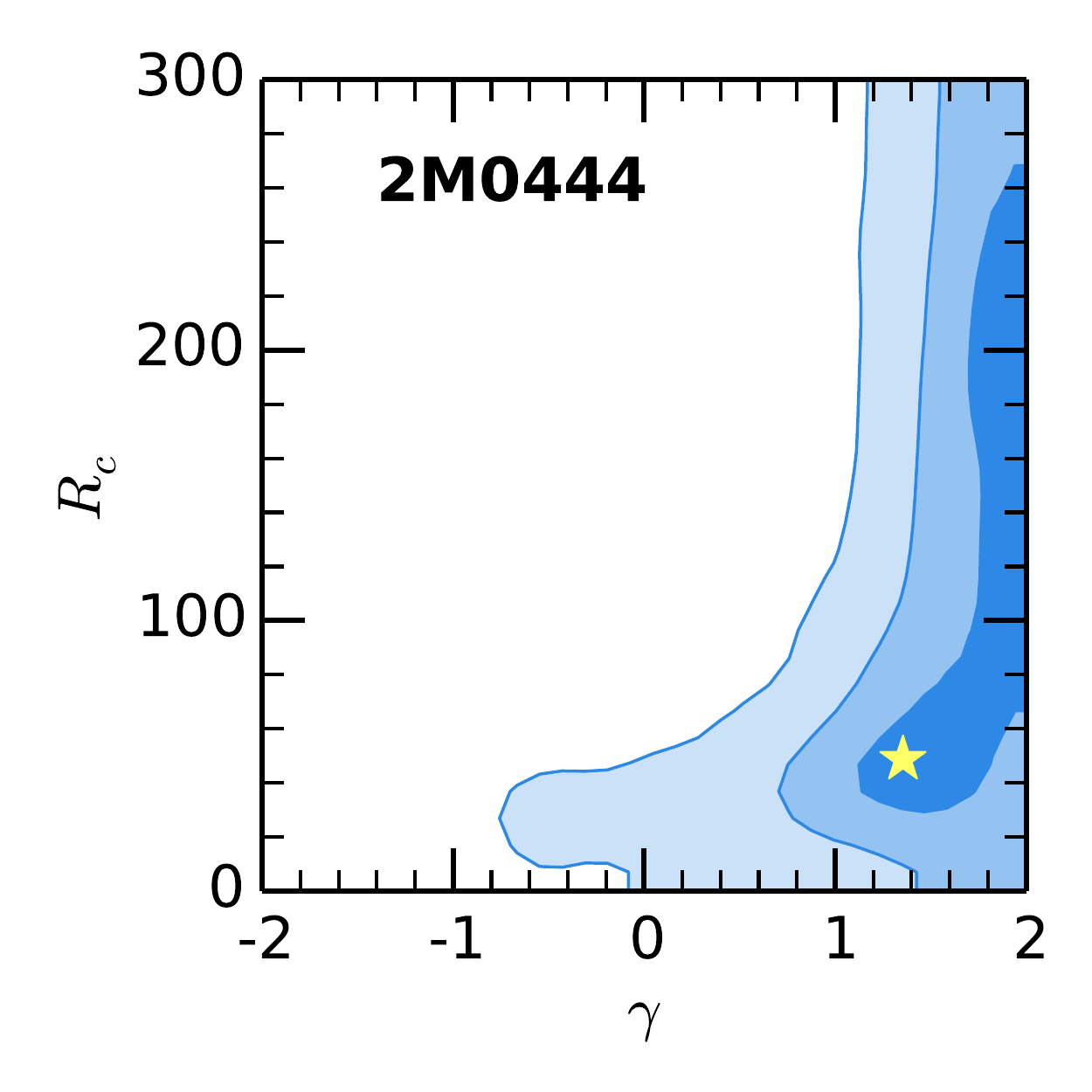}
  \includegraphics[width=4.4cm]{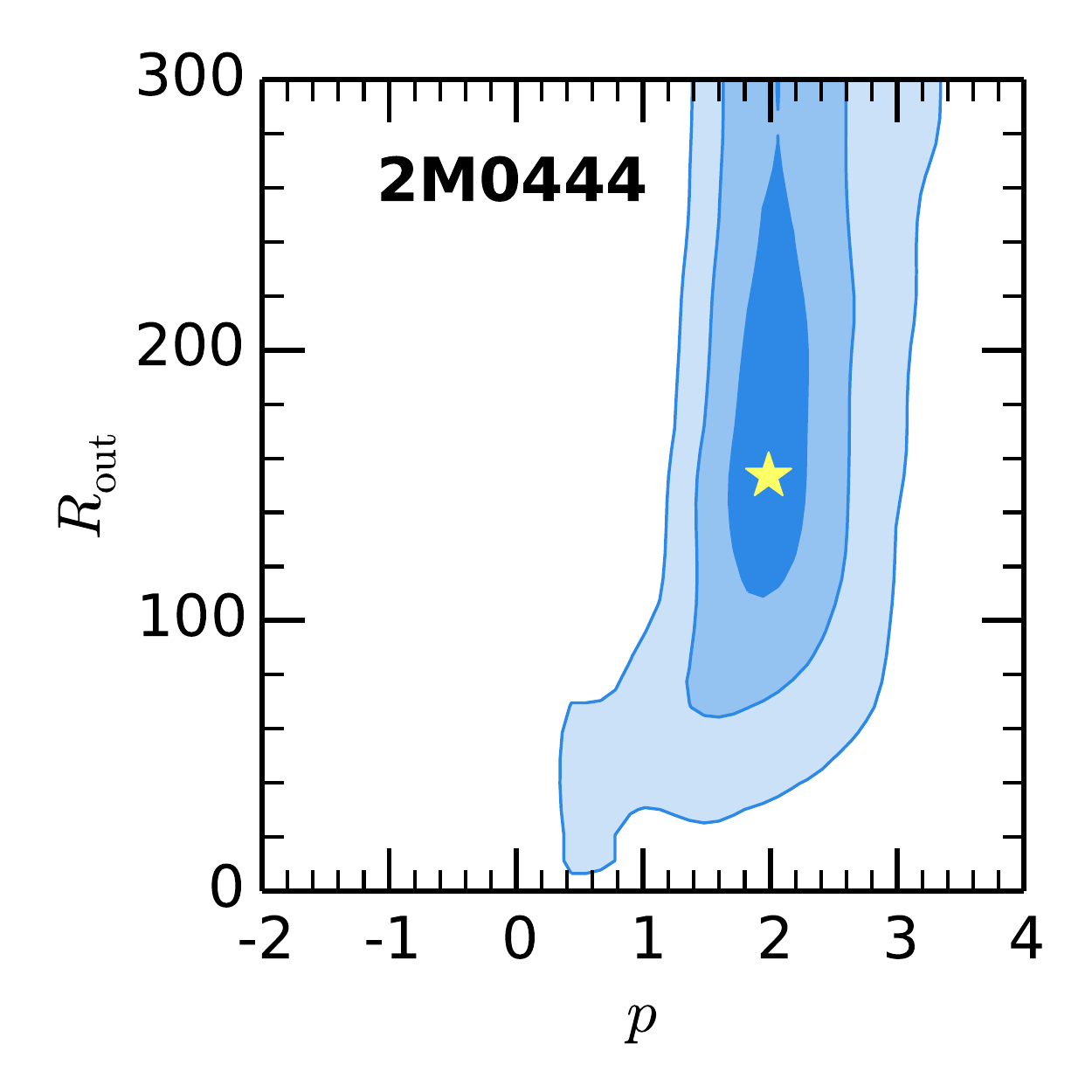}  
  \\
  \includegraphics[width=4.4cm]{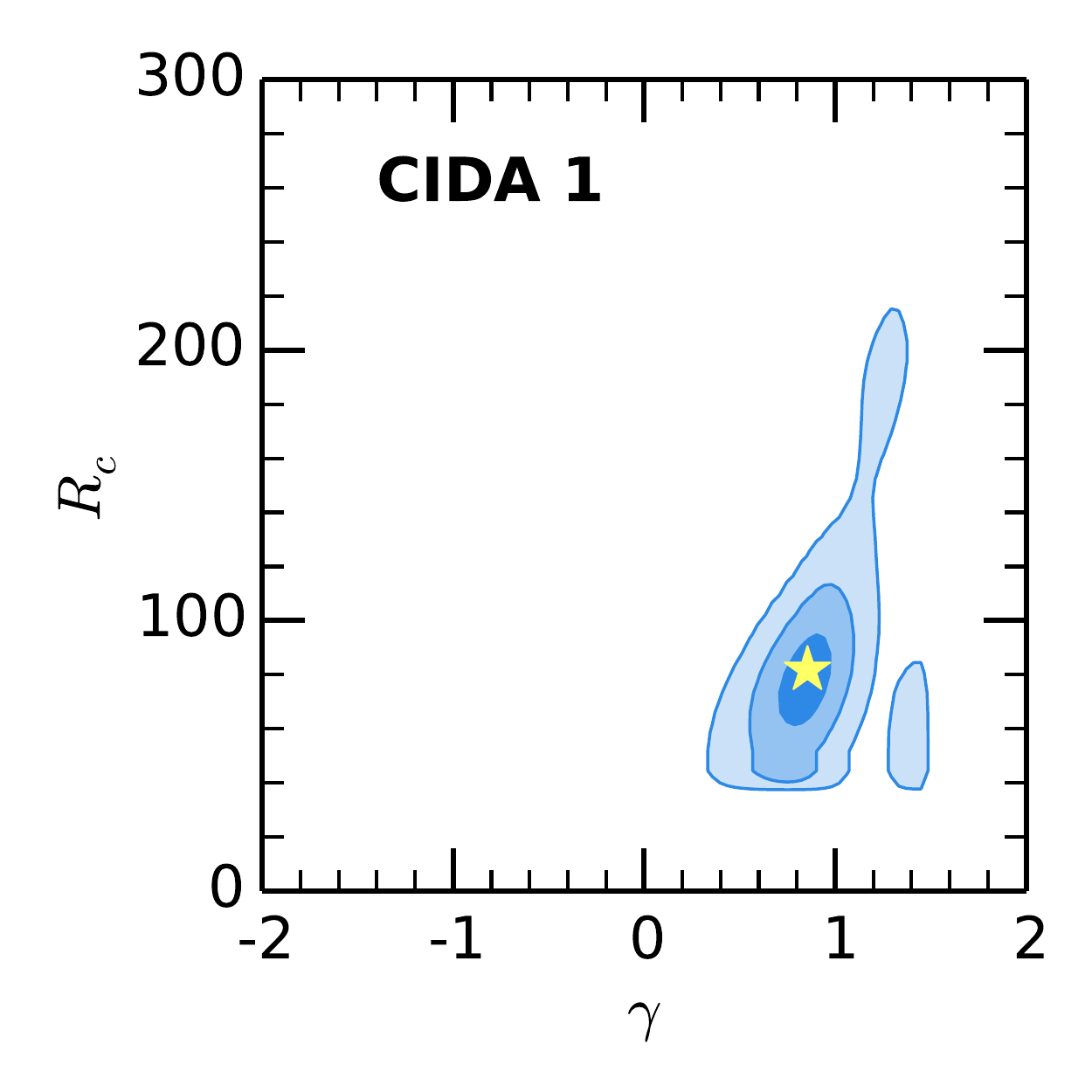}  
  \includegraphics[width=4.4cm]{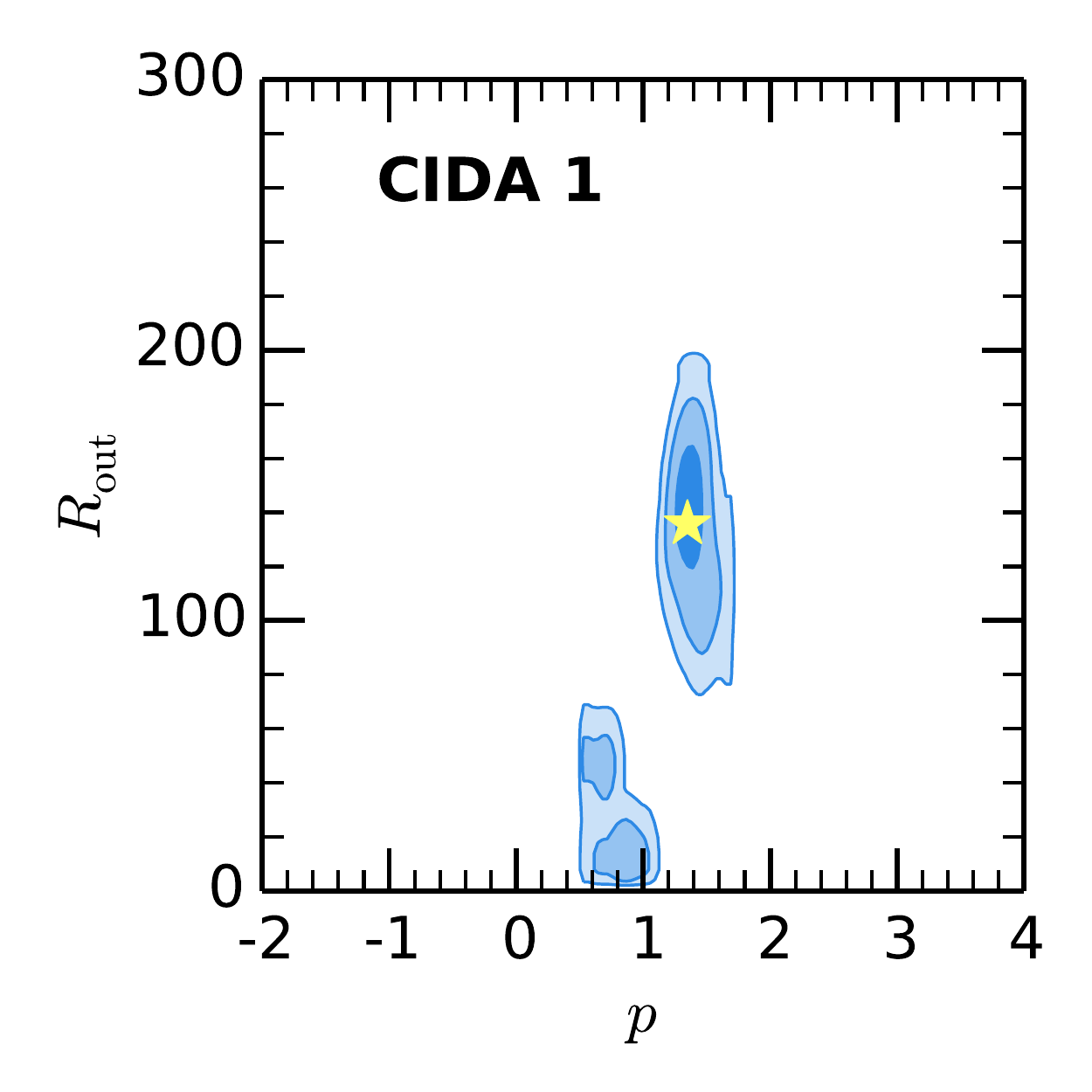}
  \\
  \includegraphics[width=4.4cm]{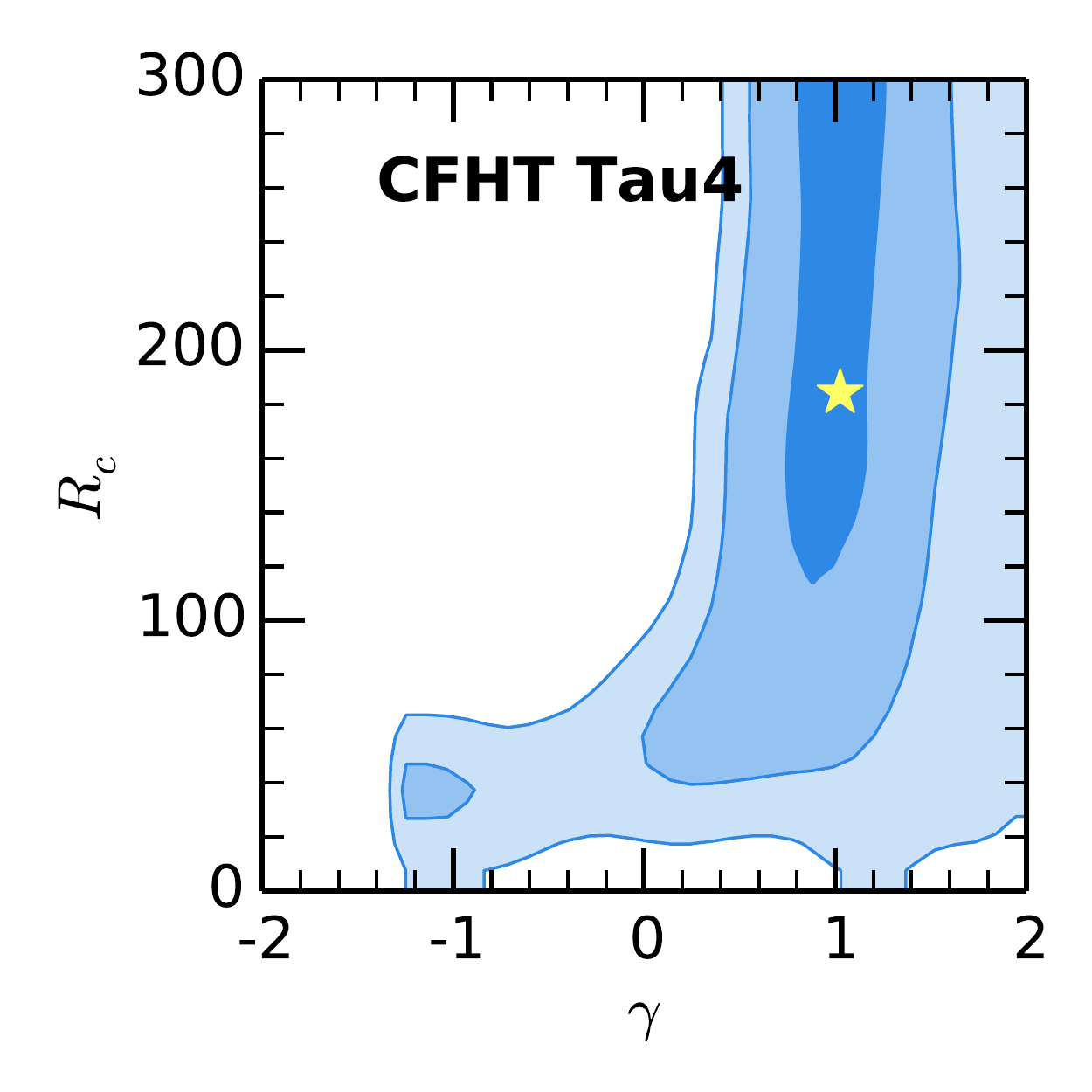}  
  \includegraphics[width=4.4cm]{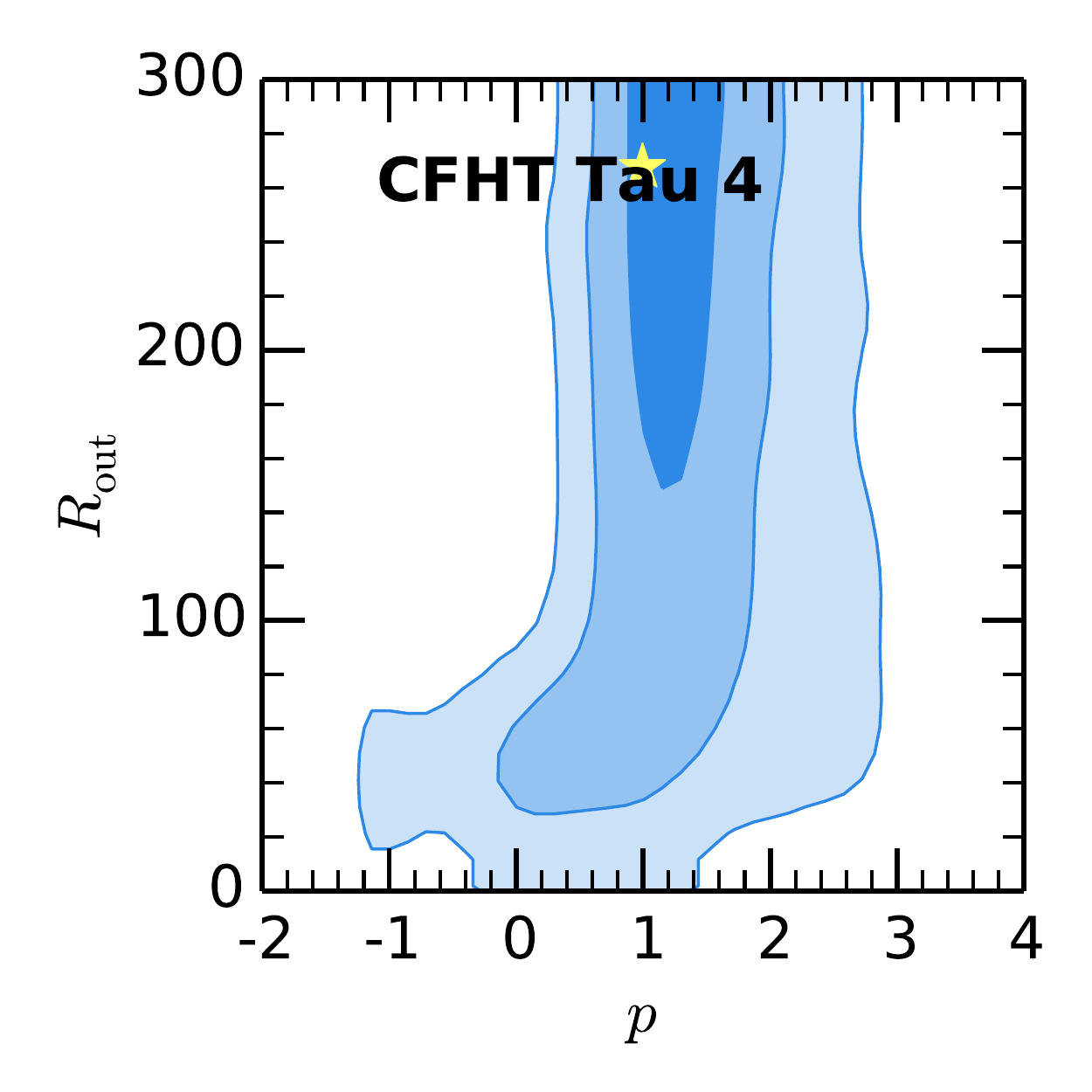}
        %\end{center}
        \caption{Two-dimensional distributions of the model parameters \{$R_c$,$\gamma$\}\ (left column) and \{$R_{out}$,$p$\}\ (right column) for the model fits of the disks in \twom\ (top row), \cida\ (middle row), and \ctfour\ (bottom row). The different shaded areas correspond to the 1, 2 and 3$\sigma$ confidence levels (from light to dark blue), as in Fig.~\ref{f_fitpar}. The yellow stars mark the parameter values for the models plotted in Fig.~\ref{f_fituvp_tau}.
    }
     \label{f_fitpar_tau}
\end{figure}

\begin{figure}
        %\begin{center}
  \includegraphics[width=6.8cm]{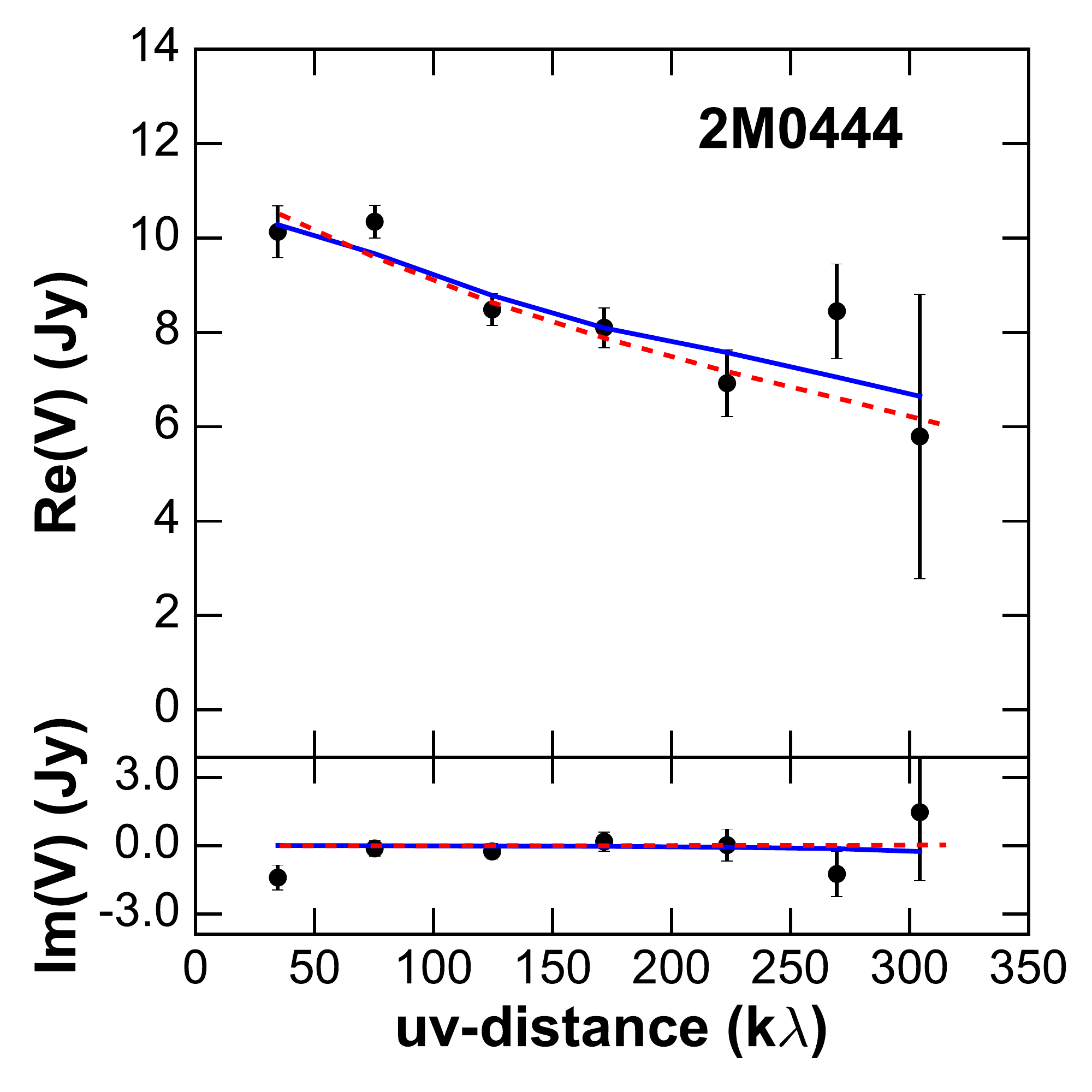}
  \\
  \includegraphics[width=6.8cm]{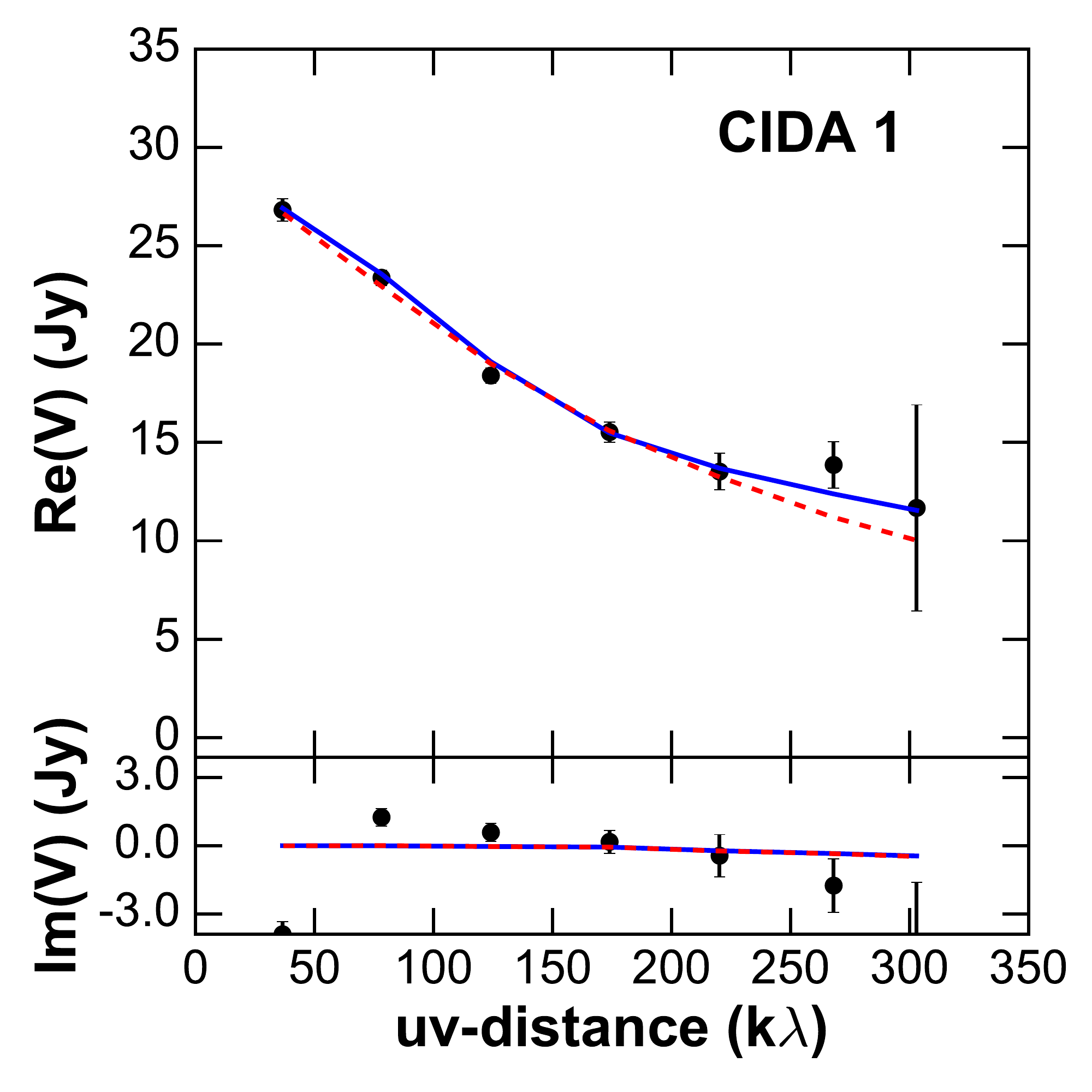}
  \\
  \includegraphics[width=6.8cm]{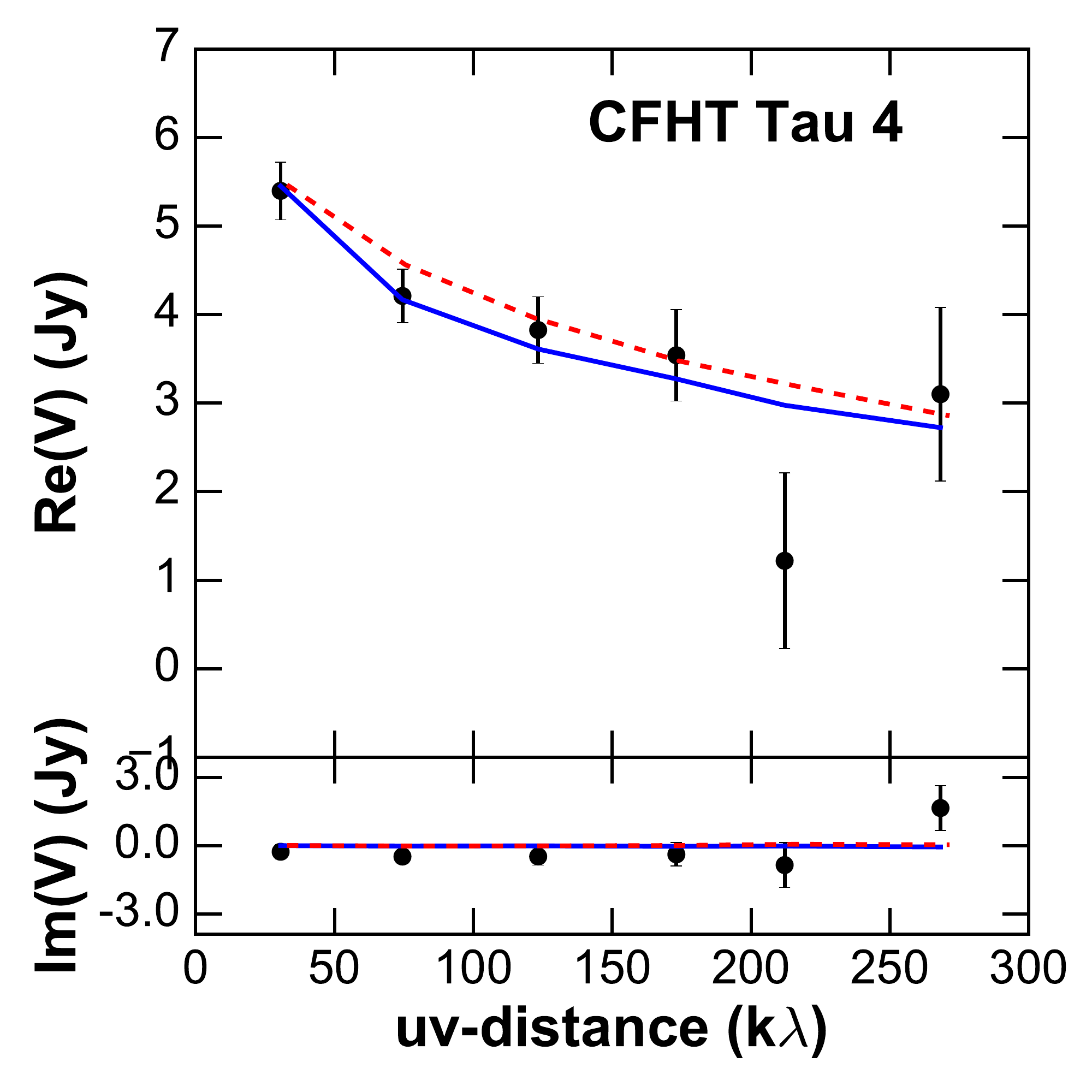}
        %\end{center}
        \caption{Comparison between observed and model visibilities for the model fits of the disks in \twom\ (top row), \cida\ (middle row), and \ctfour\ (bottom row). ALMA data is shown as black circles with errorbars, truncated power law and esponentially tapered models with blue and red lines, respectively, as in Fig.~\ref{f_fituvp}. 
    }
     \label{f_fituvp_tau}
\end{figure}
 
In this Appendix we present the results of the reanalysis of the 
Taurus BD disks ALMA 890 $\mu$m observations from \citetads{2014ApJ...791...20R}. 
We applied the procedure and parameters as described in Sect.~\ref{s_rad} to derive comparable results to our modelling of the \roph\ BDs. The results for the truncated power law fits are consistent with the results presented in \citetads{2014ApJ...791...20R}, in addition, we also performed the fits with the exponentially tapered $\Sigma(R)$. The lower values of $R_c$ for the exponentially tapered, as compared with $R_{out}$ in the truncated power law case, is consistent with the values of $\gamma$ in the range 1-1.5.
The model results are shown in graphical form in Fig.~\ref{f_fitpar_tau} and~\ref{f_fituvp_tau}, where we show the two-parameter probability distributions for the pairs \{$R_c$,$\gamma$\}\  and \{$R_{out}$,$p$\}\ (Fig.~\ref{f_fitpar_tau}), and the comparison between the ALMA observed visibilities and the visibilities of the models with the lowest $\chi^2$ in the chains (Fig.~\ref{f_fituvp_tau}).

\end{document}